\documentclass[sn-mathphys,Numbered]{sn-jnl}% Math and Physical Sciences Reference Style

\usepackage{graphicx,color}%
\usepackage{multirow}%
\usepackage{amsmath,amssymb,amsfonts,bm,pifont}%
\usepackage{amsthm}%
\usepackage{mathrsfs}%
\usepackage{mathtools}%
\usepackage[title]{appendix}%
\usepackage{xcolor}%
\usepackage{textcomp}%
\usepackage{manyfoot}%
\usepackage{booktabs}%
\usepackage{algorithm}%
\usepackage{algorithmicx}%
\usepackage{algpseudocode}%
\usepackage{listings}%
\begin{document}

\title[Article Title]{Freezing of a deformable water-saturated porous medium. Part I: THM formulation, OpenGeoSys-6 implementation and benchmarking\\ 
%\small{\color{red}(work resumed, March 4, 2026)}
}

\author*[1,2]{\fnm{Tymofiy} \sur{Gerasimov}}\email{tymofiy.gerasimov@bge-technology.de}

\author*[3]{\fnm{Christian B.} \sur{Silbermann}}\email{christian.silbermann@ifgt.tu-freiberg.de}

\author[2,3]{\fnm{Dmitri} \sur{Naumov}}\email{dmitri.naumov@ifgt.tu-freiberg.de}

\author[2,4]{\fnm{Olaf} \sur{Kolditz}}\email{olaf.kolditz@ufz.de}

\author[2]{\fnm{Haibing} \sur{Shao}}\email{haibing.shao@ufz.de}

\author[3]{\fnm{Thomas} \sur{Nagel}}\email{thomas.nagel@ifgt.tu-freiberg.de}

%%%%%%%%%%%%%%%%%%%%%%%%%%%%%%%%%%%%55
\affil*[1]{\orgdiv{Numerical Modelling Department}, \orgname{BGE Technology GmbH}, \orgaddress{\street{Eschenstraße 55}, \city{Peine}, \postcode{31224}, \country{Germany}}}

\affil*[2]{\orgdiv{Environmental Informatics Department}, \orgname{UFZ Leipzig}, \orgaddress{\street{Permoserstraße 15}, \city{Leipzig}, \postcode{04318}, \country{Germany}}}

\affil*[3]{\orgdiv{Geotechnical Institute}, \orgname{Technische Universität Bergakademie Freiberg}, \orgaddress{\street{Gustav-Zeuner-Straße 1}, \city{Freiberg}, \postcode{09599}, \country{Germany}}}

\affil*[4]{\orgdiv{Faculty of Environmental Sciences}, \orgname{Technische Universität Dresden}, \orgaddress{\street{Helmholtzstrasse 10}, \city{Dresden}, \postcode{01062}, \country{Germany}}}

\abstract{In this contribution, Part I, we present a coupled thermo-hydro-mechanical formulation for modeling and analyzing water-to-ice phase change in a deformable fully-saturated porous medium. It is implemented in the multi-physics computational platform OpenGeoSys-6. We compute a series of carefully designed benchmark problems which critically examine the corresponding formulation components and the overall implementation. Several ingredients that have a qualitative and quantitative impact on the numerical results are identified, particularly dissected and commented on. Simulations also account for soil deformation induced by freezing (as a result of 9\% volumetric expansion caused by water-to-ice phase transition).
The forthcoming Part II will complete the code verification and validation campaign by considering a real full-scale three-dimensional case study of shallow subsurface ice storage. More specifically, we simulate the ice growth process in a fully saturated soil specimen surrounding a group of borehole heat exchangers (BHEs) that contain subzero temperature coolant fluid.
A snapshot of the numerical results of this task is already depicted here in Part I as a teaser.
}

\keywords{thermo-hydro-mechanics, deformable saturated porous medium, water-to-ice phase transition, soil freezing/thawing, OpenGeoSys-6 computational platform, benchmarking, GEWS project, subsurface ice storage}



\maketitle

\tableofcontents

\begin{table}[h!]
\centering
\begin{tabular}{ r | l | l  }
	\textbf{symbol}                 & \textbf{meaning}              & \textbf{units}\\
	\hline
	$\alpha$                        & phase: solid (=S), liquid (=L), ice (=I)                   &  --\\
    \hline
	$T$                             & temperature                   & temperature, K \\
	$p$                             & pore pressure (assuming $p_\mathrm{LR}=p_\mathrm{IR}=:p$)           & pressure, Pa \\
	$\bm{u}$                        & displacement vector           & length, m \\
	\hline
	$\bm{\varepsilon}$              & strain tensor                 & -- \\
	$\bm{\sigma}^\mathrm{eff}$        & effective stress tensor     & pressure, Pa \\
	$\widetilde{\bm{w}}_\mathrm{LS}$  & Darcy filter velocity vector      & velocity, m/s \\
	\hline
	$\bm{g}$                        & gravity acceleration vector   & force/mass, N/kg \\ 
    %   &  & (length/time/time, m/s$^2$) \\
	$Q_T$                           & heat source term              & power/volume, W/m$^3$ \\
	$Q_H$                           & fluid mass (source) term      & mass/volume/time, kg/(m$^3$s) \\
	\hline
	%$\alpha_\mathrm{B}$                & Biot-Willis coefficient ($=1-\frac{K_\mathrm{S}}{K_\mathrm{SR}}\leq1$)      & -- \\
	$\phi$                          & porosity, $\phi\in(0, 1-{K_\mathrm{S}}/{K_\mathrm{SR}}]$,    & -- \\
	$\phi_\mathrm{L}$, $\phi_\mathrm{I}$  & fluid and ice fractions, resp. ($\phi_\mathrm{L}+\phi_\mathrm{I}=\phi$)           & -- \\
	\hline
	$c_{p\alpha}$                  & specific heat capacity of phase $\alpha$    & energy/mass/temperature, J/(kg K) \\
	$\varrho_{\alpha\mathrm{R}}$   & intrinsic (real) density of phase $\alpha$  & mass/volume, kg/m$^3$ \\
	$\lambda_{\alpha\mathrm{R}}$   & thermal conductivity of phase $\alpha$      & power/length/temperature, W/(m\;K) \\
    $\alpha_{T,\alpha\mathrm{R}}$  & linear thermal expansion coefficient of phase $\alpha$     & 1/temperature, 1/K \\
	$\beta_{T,\alpha\mathrm{R}}$   & volumetric thermal expansion coefficient    & 1/temperature, 1/K \\
                                   & ($\beta_{T,\alpha\mathrm{R}}=3\,\alpha_{T,\alpha\mathrm{R}}$)           &  1/temperature, 1/K \\
	$\beta_{p,\mathrm{\alpha R}}$             & compressibility of phase $\alpha$ ($\beta_{p,\mathrm{\alpha R}}=1/K_\mathrm{\alpha R}$)               & 1/stress \\
    $E_{\alpha\mathrm{R}}$, $K_{\alpha\mathrm{R}}$     & Young's and bulk moduli of phase $\alpha$ & pressure, Pa \\
	\hline
	$\mu_\mathrm{LR}$                  & liquid phase viscosity                  & pressure$\cdot$time, Pa\;s \\
	$\kappa_\mathrm{i}$                & intrinsic permeability                  & length$\cdot$length, m$^2$ \\
    $\kappa_\mathrm{L}^\mathrm{rel}$   & relative liquid permeability, $\in(0,1]$  & -- \\
    \hline
	$(\varrho c_p)^\mathrm{eff}$       & effective volumetric heat capacity       & energy/volume/temperature \\
	$\lambda^\mathrm{eff}$             & effective heat conductivity              & power/length/temperature \\
    $\varrho^\mathrm{eff}$             & effective density                        & mass/volume, kg/m$^3$ \\
    \hline
    $\ell$          & latent heat of fusion & energy/mass, J/kg  \\
    $H_\mathrm{I}$  & ice-fraction indicator function & --\\    
    $S_\mathrm{I}$  & smooth counterpart of $H_\mathrm{I}$ & -- \\
    $T_\star$       & ice-melting (=:$T_\mathrm{m}$), or water-freezing (=:$T_\mathrm{fr}$) &   \\
                    & temperature  & temperature, K \\
	\hline 
	${\bm I}$                            & second order identity tensor        & -- \\  
	$\mathbb C_{\alpha\mathrm{R}}$       & fourth order elasticity tensor      & pressure, Pa \\
	\hline
	$\Omega\subset\mathbb{R}^d$          & $d$-dimensional physical domain ($=2,3$), &   \\
	                                    & in which the THM problem is formulated &
\end{tabular}
\caption{List of symbols ({\sc NB} solid skeleton=solid matrix, solid phase=solid grain)}
\label{Nomenclature}
\end{table}

\newpage
%%%%%%%%%%%%%%%%%%%%%%%%%%%%%%%%%%%%%%%%%%%%%%%%%%%%%%%%%%%%%%%%%%%%%%%%%%%%%%%%%%%%%%%%%%%%%%%%%%%%%%%%%%%%%%%%%%%%
\section{Introduction}\label{sec1}
\begin{figure}[h!]
\begin{center}
\includegraphics[width=1.0\textwidth]{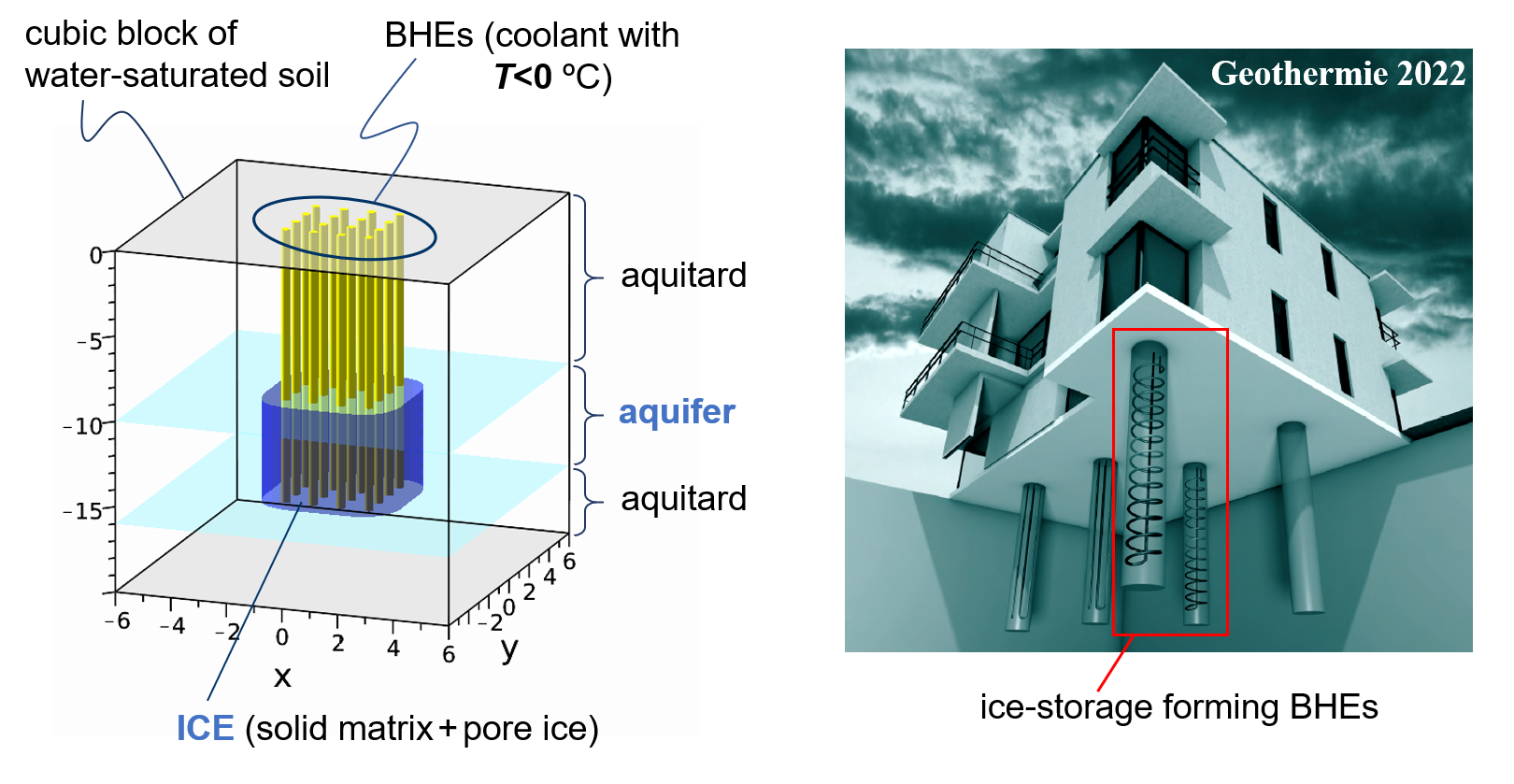}
\end{center}
\caption{Schematic setup for an underground ice-storage system and its potential application for buildings cooling (as envisioned in \cite{GPL2020}).}
\label{fig_GEWS_project}
\end{figure}

Soil freezing is a process by which pore water in a saturated porous medium undergoes a liquid-to-ice phase transition upon cooling below 0\,°C. It is a phenomenon of broad scientific and engineering significance.
In geotechnical engineering, controlled freezing of the ground has been exploited for well over a century through the technique of artificial ground freezing (AGF), wherein a network of refrigerated pipes is used to form a load-bearing ice wall that temporarily strengthens the soil and acts as a waterproof barrier during shaft sinking, tunnel construction, or deep excavation in water-bearing strata \cite{Andersland2004,Jessberger1980}.
Beyond its deliberate application in AGF, freezing and thawing of the ground occurs naturally in cold-region environments, where it governs the mechanical stability of infrastructure built on permafrost, the hydrological behavior of seasonally frozen ground, and the long-term evolution of periglacial landscapes \cite{French2007,Riseborough2008}.
More recently, subsurface freezing has emerged as a key physical process in novel shallow geothermal applications: in underground thermal energy storage (UTES) systems, groups of borehole heat exchangers (BHEs) circulating sub-zero temperature coolant are used to progressively freeze the surrounding soil, thereby storing cold energy in the form of latent heat for seasonal building cooling \cite{GPL2020}.
A representative schematic of such a ground-based energy storage system (GEWS) is illustrated in \autoref{fig_GEWS_project}.

The present work is directly motivated by the GEWS project (Geologisches Eis-W\"{a}rme-Speichersystem, BMBF grant 03G0907A-D, 2021--2024), a collaborative research initiative that developed and field-tested a patented concept for using shallow groundwater aquifers as latent-heat cold-energy stores by means of controlled, depth-targeted, cyclic freezing via BHE groups \cite{GPL2020}.
Among the project's specific objectives was the numerical prediction of freeze-thaw-induced surface deformations -- uplift and settlement -- at the GEWS demonstration site (TestUM/Wittstock), which required predictive coupled THM+freezing simulations.
This task exposed the core physical and numerical challenges of the problem: the strongly nonlinear latent-heat source term that dominates the energy balance during phase transition, the sharp reduction of hydraulic permeability as ice progressively occupies the pore space, the approximately 9\% volumetric expansion upon freezing that drives soil skeleton deformation and surface heave, and the need to represent the ice-liquid interface in a numerically stable and consistent manner across all three coupled fields.
Addressing these challenges is the central purpose of the THM+freezing formulation and OpenGeoSys-6 implementation presented in this paper.

What makes the soil freezing process particularly challenging to model is the simultaneous and tightly coupled interaction among thermal, hydraulic, and mechanical (THM) fields.
From a thermal perspective, the phase transition releases or absorbs a substantial latent heat $\ell \approx 3.34 \times 10^5$\,J/kg for water, giving rise to a highly nonlinear effective heat capacity and an internal interface between frozen and unfrozen zones, which has been mathematically addressed as the classical Stefan problem \cite{Stefan1891,Alexiades1993}.
On the hydraulic side, ice formation progressively occurs through the pore space, sharply reducing the relative permeability and thereby impeding or redirecting pore-water flow.
Meanwhile from a mechanical standpoint, the approximately 9\% volumetric expansion that accompanies the water-to-ice transition generates internal stresses and skeleton deformations. This is the physical origin of frost heave and of structural loads that BHEs and surrounding infrastructure must accommodate.
Accurately predicting these coupled phenomena in realistic three-dimensional geometries requires a consistent THM formulation that accounts for all three fields and their mutual interactions.

The theoretical foundation for coupled THM modelling of saturated porous media rests on the classical works of Biot \cite{Biot1941} and on the Theory of Porous Media (TPM), as well as the Theory of Mixtures \cite{deBoer2000,Ehlers2002}, which provide systematic continuum-mechanical frameworks for multi-phase systems.
Extension of these frameworks to include a third phase, namely ice, has been pursued by several research groups over the past two decades.
Early contributions established the thermodynamic consistency conditions and the form of the ice-fraction constitutive law, including the treatment of cryo-suction and the Clausius-Clapeyron equation that relates pore pressure to the freezing point depression \cite{Coussy2005,Nishimura2009}.
Subsequent works refined the coupled mass, momentum and energy balance equations under small deformation assumptions, incorporating the ice volume fraction as an internal variable governed by a constitutive relationship between temperature and phase saturation \cite{Bluhm2011,NaSun2017}.
Regarding hydraulic coupling, the effect of ice on permeability is commonly described through a relative-permeability function that reduces to zero as full freezing is approached, and the dependence of liquid viscosity on temperature is typically accounted for via an Arrhenius-type expression \cite{Grant2000,Zhou2014}.

From a numerical standpoint, the nonlinearity introduced by the latent-heat term, in particular by the rapid variation of the ice-fraction indicator function near the freezing point, poses well-known difficulties for iterative solvers.
Two broad strategies have been adopted in the literature: (1) sharp-interface (Stefan-problem) methods, which track the moving phase boundary explicitly, and (2) regularized (mushy-zone or smooth phase-field) methods, which smear the transition over a finite temperature interval \cite{Alexiades1993,Voller1987}.
The latter approach, adopted here, replaces the discontinuous Heaviside indicator with a smooth sigmoid function $S_\mathrm{I}(T)$ (see Eq. \eqref{Sigmoid}), whose steepness parameter $k$ controls the thickness of the numerical transition zone and, consequently, the convergence behavior of the Newton-Raphson solver.
Implementations of THM+freezing models have been reported in several finite-element codes, both commercial and academic, but comprehensive open-source implementations accompanied by systematic benchmark verification remain scarce.

Despite the maturity of the underlying theory, the following gaps motivate the present contribution.
First, while individual components of the THM+freezing system (T+F energy equation, H+F mass equation, M+F momentum equation) have each been studied and implemented separately, a unified open-source implementation that verifies all three components jointly has not been widely reported.
Second, the mechanical component in particular, including the volumetric expansion term and its effect on the skeleton stress state, is frequently simplified or neglected in existing codes, which limits their applicability to problems where frost heave and BHE deformation are of engineering interest.
Third, the emerging class of GEWS-type applications, which involvs multiple BHEs operating in concert to create a controlled frozen soil volume, demands validated forward-simulation tools capable of handling large-scale, three-dimensional, coupled THM+freezing problems.

Our motivation in this work is to extend the THM+freezing modelling feature based on OpenGeoSys-6 (OGS-6), which is a well-established, open-source multi-field computational platform for porous media \cite{Kolditz2012,Bilke2019}. As OGS-6 already supports coupled THM simulations for non-freezing scenarios, the work presented here introduces and verifies the additional constitutive ingredients required to activate the freezing/thawing process within this framework, culminating in the GEWS-motivated application presented in the companion paper (Part II).

This paper is organized as follows.
Section~\ref{sec:THMformulation} presents the THM+freezing (THM+F) formulation: the ice-fraction indicator function and its regularized sigmoid counterpart are introduced first, followed by the full coupled system of partial differential equations governing the T+freezing, H+freezing, and M+freezing sub-problems.
Section~\ref{sec:Benchmarks} documents a series of carefully designed benchmark problems that examine each component of the formulation and the corresponding OGS-6 implementation. This includes a one-dimensional Stefan melting problem with analytical solution, a two-dimensional verification using the method of manufactured solutions, and a soil freezing process around a BHE cross-section compared against an independent FreeFem++ implementation.
Several numerical ingredients that have a qualitative and quantitative impact on the results, in particular the choice of sigmoid steepness parameter $k$ and the convexity of the underlying energy functional, are identified and analyzed as well, with its details provided in the Appendix.
% HS: I am not sure what is the intention of Part II. The following text is just my speculation. Feel free to modify it.
Section~\ref{sec:summary} summarizes the results and provides an outlook. A brief preview of the full-scale GEWS application (30-day cooling simulation around a group of BHEs in three dimensions) is included as a teaser for Part~II.

%%%%%%%%%%%%%%%%%%%%%%%%%%%%%%%%%%%%%%%%%%%%%%%%%%%%%%%%%%%%%%%%%%%%%%%%%%%%%%%%%%%%%%%%%%%%%%%%%%%%%%%%%%%%%%%%%%%%
\section{THM formulation with liquid-to-ice phase change}
\label{sec:THMformulation}
This section presents an extension of a classical THM formulation for deformable saturated porous medium to the one that accounts for liquid-to-ice phase change. In the following, we term this the THM+freezing (or simply THM+F) model, though the inverse process, namely, ice-thawing is naturally captured with this formulation as well. The so-called ice-fraction indicator function, one of the key formulation ingredients, is first detailed. The coupled system of the corresponding equations then follows. The related T+freezing, H+freezing and M+freezing parts of the system are presented without derivation details (we only refer to the series works \cite[and references therein]{Bluhm2011}).

\subsection{The ice fraction indicator function} 
Below, we briefly present one of the pillar ingredients of the THM+freezing formulation, the so-called ice fraction indicator function and its regularized (smooth) counterpart we require for numerical simulation. 

Let $H$ be a standard Heaviside function, that is, $H(a)=1$, if $a\geq0$ and $H(a)=0$, otherwise. Consider the function
\begin{equation*}
H_\mathrm{I}(T):=1-H(T-T_\star)=
\left\{
\begin{tabular}{ll}
$1$, & $T< T_\star$,   \\  [0.2cm]
$0$, & $T\geq T_\star$,
\end{tabular}
\right.
\end{equation*}
where $T=T({\bm x},t)$ is a temperature variable and $T_\star:=0$ °C ($273.15$ K) denotes -- depending on the process ``direction'' -- either ice melting temperature $T_\mathrm{m}$, or water freezing temperature $T_\mathrm{fr}$. In the following, due to our earlier introduced notion of the ``THM+freezing'' model, we opt for using  the notation $T_\star:=T_\mathrm{fr}$. Function $H_\mathrm{I}$ aims at distinguishing between the liquid and the ice phase of the fluid (values 0 and 1, resp.) within the physical domain $\Omega$ at any instant of time. It is, hence, called ice-phase (ice fraction) {\em indicator} function.

\begin{figure}[h!]
\begin{center}
\includegraphics[width=1.0\textwidth]{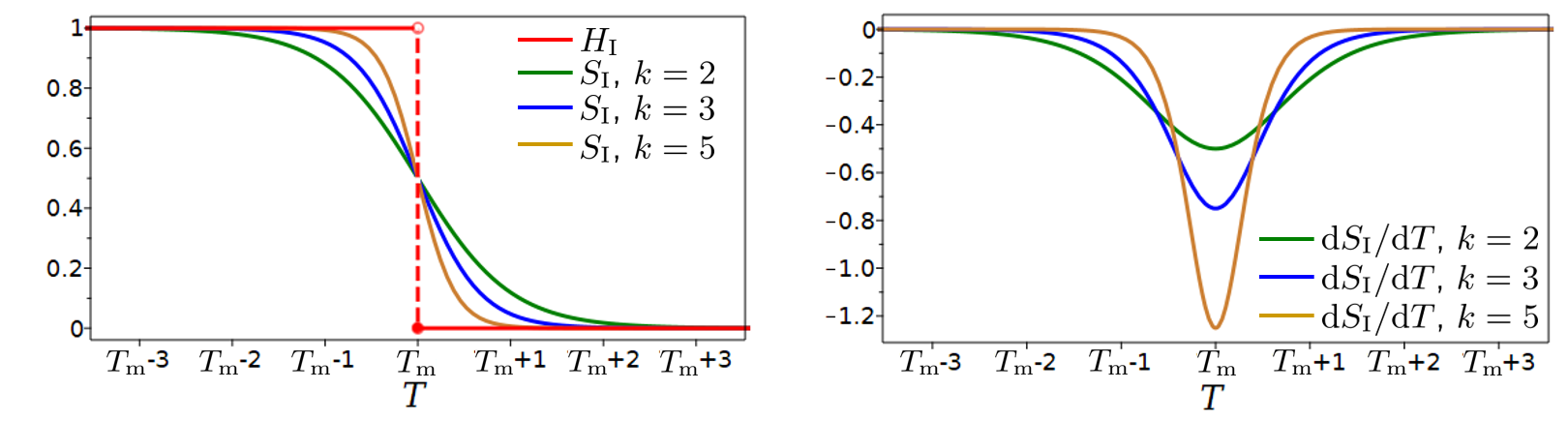}
\end{center}
\caption{On the left: plots of the ice-fraction indicator function $H_\mathrm{I}$ and its regularized counterpart $S_\mathrm{I}$; on the right: plots of the first-order derivative of $S_\text{I}$.}
\label{fig_Sigm_f}
\end{figure}

For computational purposes, its regularized (that is, a smooth/infinitely differentiable) counterpart is typically used:
\begin{equation}
    S_\mathrm{I}(T):=\frac{1}{1+e^{k(T-T_\mathrm{fr})}},\quad k>0.
\label{Sigmoid}
\end{equation}
It is a sigmoid-shaped function, see \autoref{fig_Sigm_f}, left, where $S_\text{I}$ is depicted for various values of $k>1$ along with the function $H_\mathrm{I}$. Parameter $k>0$ governs the steepness of $S_\mathrm{I}$, that is, the thickness of a transition zone between the two states and, as a result, the thickness of the localization zone of $\frac{\mathrm{d} S_\text{I}}{\mathrm{d} T}$ (explicitly, in terms of the temperature variable, and implicitly, in space), see \autoref{fig_Sigm_f}, right. Thus, in a regularized manner, the ice fraction in $\Omega$ at any moment of time $t\geq0$ is defined as $\phi_\mathrm{I}:=\phi_\mathrm{I}(T)=\phi S_\mathrm{I}(T)$. As a result, in the following, we also have $\frac{\mathrm{d}\phi_\mathrm{I}}{\mathrm{d} T}=\phi\frac{\mathrm{d} S_\mathrm{I}}{\mathrm{d} T}$ for constant porosity $\phi$.

Some hints on how to optimally (in a certain sense) choose $k$ is proposed and discussed in the appendix.

%%%%%%%%%%%%%%%%%%%%%%%%%%%%%%%%%%%%%%%%%%%%%%
\subsection{Coupled system of PDEs characterizing THM process with liquid-to-ice phase-change (THM+freezing)}

This section presents the coupled system of THM equations, which already contains a freezing constitutive component/ingredient. As already mentioned in the introductory part, the system presented herein has not been ``copy-pasted" from one of the available reference literature sources. Rather, our goal was to implement water freezing / ice melting into the already existing THM formulation within OGS in order to enable soil freezing modeling. As code developers, we revisited one of the most comprehensive descriptions available, e.\,g. in \citep{Bluhm2011}, and made our efforts -- following the standard derivation procedures -- first, to re-derive the formulation. This aimed not only to exclude potential mistakes (typos) in the published results, but actually to get familiar with constitutive machinery behind formulations accounting for a phase-change. Such kind of expertise is particularly valuable once there is a need of further modification and advancement of the model and/or the implementation. Re-derivation appeared by no means a straightforward task (already technically), given that the original coupled THM model -- containing only the standard solid and liquid phase -- is cumbersome enough. Introduction of an extra phase (such as ice), more precisely, of a liquid-to-ice phase-transition leads to extra complexity requiring non-trivial modifications of all separate (T-, H- and M-) equations. 

We ended up with what is identical to the published formulations (in the corresponding limit of a small deformation assumption), what also naturally reduces to the classical formulation in a limiting case of $\phi_\mathrm{I}\rightarrow 0$, and, finally, what yields reasonable modeling results for multiple carefully-designed benchmarks with evident phenomenological outcome (cf. \autoref{sec:Benchmarks}).

Remarks on nomenclature and assumptions during the derivation procedure:

\begin{itemize}
\item  material time-derivatives for a quantity  $A\in\{T,p,S_\mathrm{I},...\}$: $(A)^\prime_\mathrm{S}:=\frac{\partial A}{\partial t}+\nabla A\cdot \bm v^\mathrm{S}$ with $\bm v^\mathrm{S}=(\bm u^\mathrm{S})^\prime_\mathrm{S}$ being solid skeleton/matrix velocity. Assuming $\bm v^\mathrm{S}$ negligible, one arrives at and use $(A)^\prime_\mathrm{S}:=\frac{\partial A}{\partial t}$,

\item we set the Biot-Willis coefficient $\alpha_\mathrm{B}:=1-\frac{K_\mathrm{S}}{K_\mathrm{SR}}$ to 1,
   
\item the simplifying assumption $p_\mathrm{SR}=p_\mathrm{LR}=p_\mathrm{IR}=:p$ has been used\footnote{More generally, one may have
\begin{equation*}
    p_\mathrm{IR}=\frac{\varrho_\mathrm{IR}}{\varrho_\mathrm{LR}}p_\mathrm{LR}-\varrho_\mathrm{IR}\ell\ln{\left(\frac{T}{T_\mathrm{fr}}\right)},
%\label{ClCl}
\end{equation*}
see, e.g., \cite{Nishimura2009,NaSun2017}. Using this and the related derivative $(p_\mathrm{IR})^\prime_\mathrm{S}$, the resulting mass balance equation (\ref{BalMass}) is expected to be computationally rather involved.},

\item pore ice-fraction definition is $\phi_\mathrm{I}:=\phi_\mathrm{I}(T)=\phi S_\mathrm{I}(T)$, such that pore liquid fraction reads $\phi_\mathrm{L}=\phi-\phi_\mathrm{I}=\phi\left(1-S_\mathrm{I}(T)\right)$; when it seems appropriate to emphasize the corresponding fractions, we use the notations $\phi_\mathrm{I}$ and $\phi_\mathrm{L}$ explicitly, otherwise, it will be rather $\phi S_\mathrm{I}(T)$ and $\phi\left(1-S_\mathrm{I}(T)\right)$, resp.; notice that $\phi_\mathrm{S}:=1-\phi$ is never used explicitly, instead, it is always the term $1-\phi$ which is present in the formulations.
\end{itemize}

%%%%%%%%%%%%%%%%%%%%%%%%%%%%%%%%%%%%%%%%%%5
\textbf{Energy balance equation with phase change (T+freezing):}
\begin{equation}
    \left(
    (\varrho c_p)^\mathrm{eff}-\ell\varrho_\mathrm{IR}\frac{\mathrm{d}\phi_\mathrm{I}}{\mathrm{d} T}
    \right)(T)^\prime_\mathrm{S}
    -\mathrm{div}
    \left(\lambda^\mathrm{eff} \nabla T\right)
    +\varrho_\mathrm{LR}c_{p\mathrm{L}}
    \nabla T\cdot\widetilde{\bm w}^\mathrm{LS}=Q_T \quad\mbox{in}\;\Omega,
\label{BalEn}
\end{equation}
where the effective quantities read
\begin{equation*}
    (\varrho c_p)^\mathrm{eff}:=
    (1-\phi)\varrho_\mathrm{SR}c_{p\mathrm{S}}+(\phi-\phi_\mathrm{I}(T))\varrho_\mathrm{LR}c_{p\mathrm{L}}+\phi_\mathrm{I}(T)\varrho_\mathrm{IR}c_{p\mathrm{I}},
\end{equation*}
\begin{equation*}
    \lambda^\mathrm{eff}:=
    (1-\phi)\lambda_\mathrm{SR}+(\phi-\phi_\mathrm{I}(T))\lambda_\mathrm{LR}+\phi_\mathrm{I}(T)\lambda_\mathrm{IR},
\end{equation*}
and $\ell=3.34\cdot 10^{5}$ J/kg is the enthalpy called latent heat of fusion. The liquid filter velocity $\widetilde{\bm w}^\mathrm{LS}$ is assumed to be subject to Darcy's law: 
\begin{equation}
\widetilde{\bm w}^\mathrm{LS}
:=\phi_\mathrm{L}\bm w^\mathrm{LS}
=\phi_\mathrm{L}\bm (\bm v^\mathrm{L}- \bm v^\mathrm{S})
=\kappa_\mathrm{L}^\mathrm{rel}\frac{\kappa_\mathrm{i}\bm I}{\mu_\mathrm{LR}}\cdot
(-\nabla p+\varrho_\mathrm{LR}\bm g),
\label{Darcy}
\end{equation}
where $\kappa_\mathrm{L}^\mathrm{rel}$ and $\kappa_\mathrm{i}$ are relative liquid phase permeability and porous medium intrinsic permeability, respectively. In order to provide the decrease of liquid water flow in the region where ice forms, both relative permeability and viscosity should depend on temperature (see \autoref{subsec:BenchmarkH+freezing}).

%%%%%%%%%%%%%%%%%%%%%%%%%%%%%%%%%%%%%%%%%%%%%%%%%%5
\textbf{Mass balance equation with phase change (H+freezing):}
\begin{equation*}
\biggl\{
\frac{1-\phi}{K_\mathrm{SR}}\Bigl[
\varrho_\mathrm{LR}(1-S_\mathrm{I}(T))+\varrho_\mathrm{IR}S_\mathrm{I}(T)\Bigr]
+\frac{\phi}{K_\mathrm{LR}}\varrho_\mathrm{LR}(1-S_\mathrm{I}(T))
\biggr.
\end{equation*}
\begin{equation*}
\biggl.+\frac{\phi}{K_\mathrm{IR}}\varrho_\mathrm{IR}S_\mathrm{I}(T)
\biggr\}(p)^\prime_\mathrm{S}
\end{equation*}
\begin{equation*}
-\biggl\{ (1-\phi)\beta_{T,\mathrm{SR}}
\Bigl[\varrho_\mathrm{LR}(1-S_\mathrm{I}(T))+\varrho_\mathrm{IR}S_\mathrm{I}(T)\Bigr]
+\phi\beta_{T,\mathrm{LR}}\varrho_\mathrm{LR}(1-S_\mathrm{I}(T))
\biggr.
\end{equation*}
\begin{equation*}
\biggl. +\phi\beta_{T,\mathrm{IR}}\varrho_\mathrm{IR}S_\mathrm{I}(T)
\biggr\}(T)^\prime_\mathrm{S}
\end{equation*}
\begin{equation*}
+\Bigl[\varrho_\mathrm{LR}(1-S_\mathrm{I}(T))+\varrho_\mathrm{IR}S_\mathrm{I}(T)\Bigr]
\mathrm{div}\left(\bm v^\mathrm{S}\right)
+\mathrm{div}
\left(
\varrho_\mathrm{LR}\widetilde{\bm w}^\mathrm{LS}
\right)
\end{equation*}
\begin{equation}
-\left(\varrho_\mathrm{LR}-\varrho_\mathrm{IR}\right)
(\phi_\mathrm{I})^\prime_\mathrm{S}=Q_H \quad\mbox{in}\;\Omega,
\label{BalMass}
\end{equation}
where the phase bulk stiffness is the inverse of the corresponding compressibility and the divergence of the solid phase velocity equals the volumetric strain rate involved in the solid motion. Note that the term followed by $(\phi_\mathrm{I})^\prime_\mathrm{S} \equiv \phi\, (S_\mathrm{I})^\prime_\mathrm{S}$ acts as a localization term reflecting the interface and coupling between the phases. It is only active during the phase transition, otherwise it disappears.

%%%%%%%%%%%%%%%%%%%%%%%%%%%%%%%%%%%%%%%%%%%%%%%%%%
\textbf{Momentum balance equation with phase change (M+freezing):}
The general form of the balance equation of momentum for the total mixture reads
\begin{equation*}
    \mathrm{div}\Bigl[\bm\sigma_\mathrm{SI}^\mathrm{eff}
    -\bigl(S_\mathrm{I}(T)p_\mathrm{IR}+(1-S_\mathrm{I}(T))p_\mathrm{LR}\bigr){\bm I}\Bigr]
    +\varrho^\mathrm{eff}{\bm g}={\bm 0}\quad\mbox{in}\;\Omega,
\end{equation*}
where $\bm\sigma_\mathrm{SI}^\mathrm{eff}$ is the so-called effective stress of the solid matrix--ice mixture, $p_\mathrm{LR}$ and $p_\mathrm{IR}$ are pore liquid- and ice-phase pressure, respectively. Finally,
\begin{equation*}
\varrho^\mathrm{eff}=(1-\phi)\varrho_\mathrm{SR}+(\phi-\phi_\mathrm{I}(T))\varrho_\mathrm{LR}+\phi_\mathrm{I}(T)\varrho_\mathrm{IR},
\end{equation*}
is the effective mixture density. Before we elaborate the expression for the effective stress, notice that the assumption of $p_\mathrm{LR}=p_\mathrm{IR}=:p$ we agreed on earlier, the momentum balance equation we arrive at and also use in our further analysis reads
\begin{equation}
    \mathrm{div}\bigl[\bm\sigma_\mathrm{SI}^\mathrm{eff}
    -p{\bm I}\bigr]
    +\varrho^\mathrm{eff}{\bm g}={\bm 0}\quad\mbox{in}\;\Omega.
\label{BalMom}
\end{equation}

The {\em original} representation of the effective stress in (\ref{BalMom}) reads
\begin{align}
    \bm\sigma_\mathrm{SI}^\mathrm{eff}(\bm\varepsilon_\mathrm{S},\bm\varepsilon_\mathrm{I},T)
    &= \bm\sigma_\mathrm{S} + \phi_\mathrm{I} \bm\sigma_\mathrm{IR} = 
    \mathbb{C}_\mathrm{S}:\bigl(\bm\varepsilon_\mathrm{S}-\alpha_{T,\mathrm{SR}} (T-T_0){\bm I}\bigr) \nonumber \\ 
    &+ \phi_\mathrm{I}(T) \mathbb{C}_\mathrm{IR}:\bigl(\bm\varepsilon_\mathrm{I}-\alpha_{T,\mathrm{IR}} (T-T_\mathrm{fr}){\bm I} - \alpha_{\phi_\mathrm{I}} (S_\mathrm{I}(T)-S_{\mathrm{I}0}){\bm I} \bigr).
\label{sigESI_2}
\end{align}
We derived this following the variation of the Helmholtz energies for the solid skeleton/matrix and ice fraction proposed in \cite{Bluhm2011}, as well as using, in our case, a small deformation assumption. We term it the {\em original} since, in the present form, it is seemingly not ready to be used in the computations, but yet requires explanation and, possibly, further elaboration. 

Indeed, there are ingredients in (\ref{sigESI_2}), whose meaning is already clear. For example, the parameter $\alpha_{\phi_\mathrm{I}}$ referred to as the linear expansion coefficient due to water-to-ice phase change. During the transition, the volume of freezing water expands by 9 percent, thus resulting in $\alpha_{\phi_\mathrm{I}}=\frac{1}{3}\ln(1.09)\approx0.03$. Next to this, the coefficient $S_{\mathrm{I}0}$ stands for the ice fraction if it is initially present in the domain $\Omega$, $T_{\mathrm{S}0}$ is the reference temperature for the solid matrix material, and $T_\mathrm{fr}$ is the freezing temperature already present in the Sigmoid function.

One can notice, however, that $\bm\sigma_\mathrm{SI}^\mathrm{eff}$ also  contains two unknowns, the solid matrix strain $\bm\varepsilon_\mathrm{S}$ and the ice-fraction strain $\bm\varepsilon_\mathrm{I}$. According to Bluhm et al. \cite[section 3]{Bluhm2011}, these seem to be dependent quantities:
\begin{equation}
\bm\varepsilon_\mathrm{S}=\bm\varepsilon_{\mathrm{S}0}+\bm\varepsilon_\mathrm{I},
\label{impt}
\end{equation} 
where $\bm\varepsilon_{\mathrm{S}0}$ represents the strain accrued by the solid matrix/skeleton before the onset of the water-to-ice phase transition. It is also mentioned there that ``motions of ice and solid are {\em identical} except at an initial solid motion'', what -- to our understanding -- means that the deformed configuration of the solid matrix at ${\bm x}\in\Omega$ {\em prior} to ice formation must be 'seen' by the ice as a reference configuration.

\begin{figure}[h!]
\begin{center}
\includegraphics[width=1.0\textwidth]{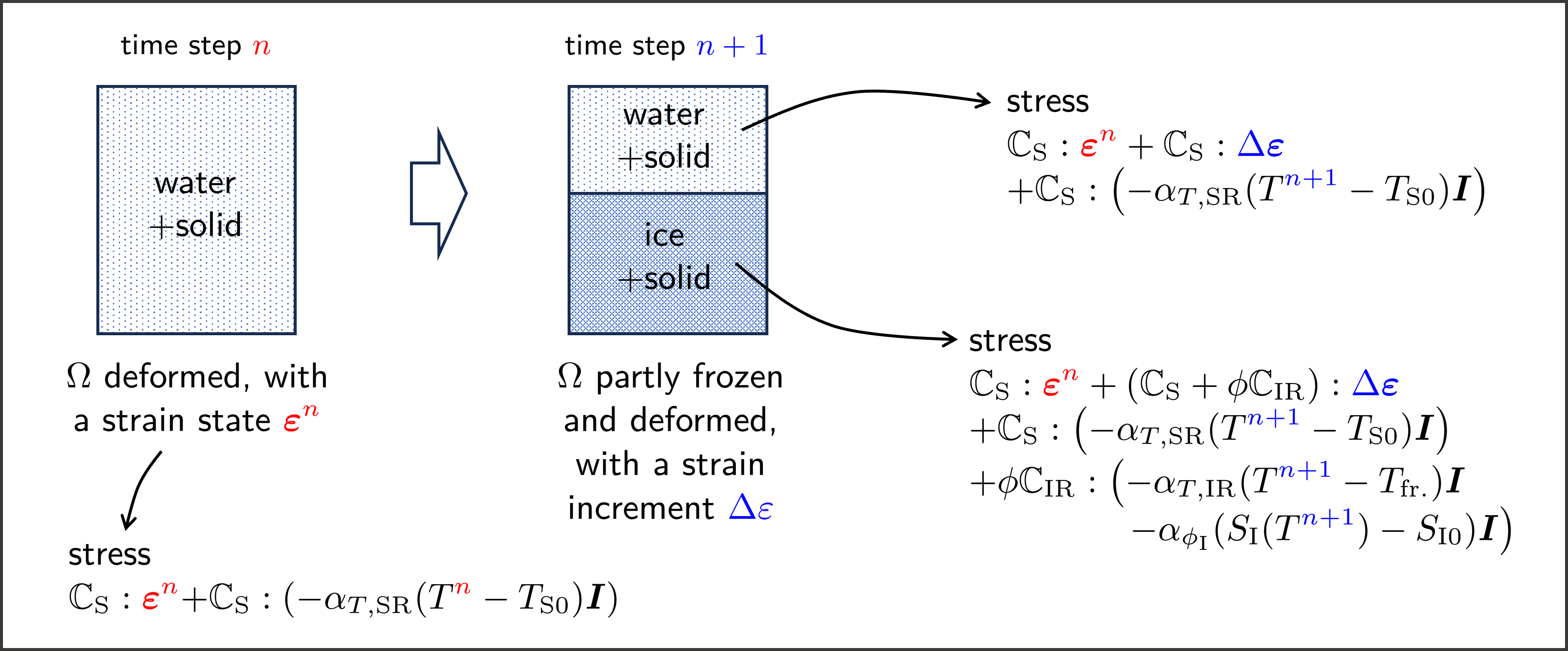}
\end{center}
\caption{Our incremental stress--strain relation (schematically), which is capable to account for liquid-to-ice phase transition. The split of the corresponding stresses into mechanical and thermal contributions are emphasized to trace the difference between the steps more explicitly.}
\label{fig_00}
\end{figure}

Technically, by plugging (\ref{impt}) into (\ref{sigESI_2}), one arrives at the relation
\begin{align}
    \bm\sigma_\mathrm{SI}^\mathrm{eff}(\bm\varepsilon_\mathrm{S},T)
    &= \mathbb{C}_\mathrm{S}:\bigl(\bm\varepsilon_\mathrm{S}-\alpha_{T,\mathrm{SR}} (T-T_0){\bm I}\bigr) \nonumber \\ 
    &+ \phi_\mathrm{I}(T) \mathbb{C}_\mathrm{IR}:\bigl(\bm\varepsilon_\mathrm{S}-\bm\varepsilon_{\mathrm{S}0}-\alpha_{T,\mathrm{IR}} (T-T_\mathrm{fr}){\bm I} - \alpha_{\phi_\mathrm{I}} (S_\mathrm{I}(T)-S_{\mathrm{I}0}){\bm I} \bigr).
\label{sigESI_3}
\end{align}
In the next section, we present our (re-)interpretation of (\ref{sigESI_3}) with more explicit identification of the corresponding $\bm\varepsilon_\mathrm{S}$ and $\bm\varepsilon_{\mathrm{S}0}$. To this end, we use our notations, as well as the incremental form of a strain-stress relationship accounting for water-to-ice phase change. Our idea is briefly illustrated in Figure \ref{fig_00}. After the following explanation, exactly this version of (\ref{sigESI_3}) will be implemented in the OGS code and verified.

%%%%%%%%%%%%%%%%%%%%%%%%
%%%%%%%%%%%%%%%%%%%%%%%

\subsection{Our version of the M+freezing formulation based on Eq.~(\ref{sigESI_3})}

Suppose that at a time step $n$, we have a fully saturated porous medium specimen (domain $\Omega$), whose temperature $T^n$ is higher than the freezing one and whose strain state is given by $\bm\varepsilon^n$. No ice-phase is yet present in $\Omega$ (this is termed ``water+solid'' in the related graph of Figure \ref{fig_00}), so $\bm\varepsilon^n$ is the strain of the solid matrix. The linearly elastic stress state has the following standard representation:
\begin{equation*}
    \bm\sigma^n:= \mathbb{C}_\mathrm{S}:\bigl(\bm\varepsilon^n-\alpha_{T,\mathrm{SR}} (T^n-T_{\mathrm{S}0}){\bm I}\bigr).
\end{equation*}

Let, at a time step $n+1$, the specimen state be characterized in a way that it has been partially frozen and both parts (``ice+solid'' and ``water+solid'', Figure \ref{fig_00}) have also been further deformed. The previous strain state $\bm\varepsilon^n$ can be viewed in this case as a reference configuration for both the already frozen-deformed and the unfrozen-deformed subdomains. The major difference between these subdomains in terms of the stress representation lies in the stiffness tensor to which the strain increment $\Delta\bm\varepsilon$ is associated -- it is $\mathbb{C}_\mathrm{S}$ for the solid matrix and $\mathbb{C}_\mathrm{S}+\phi\mathbb{C}_\mathrm{IR}$ for the mixture of solid matrix and ice-filled pores. Further, the thermal part of the ``ice+solid'' stress tensor contains the additional contribution that enables volumetric expansion due to water-to-ice phase change, namely, the term $\alpha_{\phi_\mathrm{I}} (S_\mathrm{I}(T)-S_{\mathrm{I}0}){\bm I}$.

We can collapse the related ``ice+solid'' and ``water+solid'' expressions for stress at time step $n+1$ into one equation valid in the entire domain $\Omega$. This reads:
\begin{align*}
    \bm\sigma^{n+1}
    &:= \mathbb{C}_\mathrm{S}:\bigl(\bm\varepsilon^n +\Delta\bm\varepsilon -\alpha_{T,\mathrm{SR}} (T^{n+1}-T_{\mathrm{S}0}){\bm I}\bigr) \nonumber \\ 
    &+ \phi_\mathrm{I}(T) \mathbb{C}_\mathrm{IR}:\left(\Delta\bm\varepsilon - \alpha_{T,\mathrm{IR}} (T^{n+1}-T_\mathrm{fr}){\bm I} - \alpha_{\phi_\mathrm{I}} (S_\mathrm{I}(T^{n+1})-S_{\mathrm{I}0}){\bm I} \right),
%\label{Int2}
\end{align*}
where $\bm\varepsilon^n$ is assumed to be the input known from the previous time step and $\Delta\bm\varepsilon$ is the unknown increment to be solved for. Using the notation $\bm\varepsilon^{n+1}:=\bm\varepsilon^n +\Delta\bm\varepsilon$, the above equation is rewritten as follows
\begin{align}
    \bm\sigma^{n+1}
    &:= \mathbb{C}_\mathrm{S}:\bigl(\bm\varepsilon^{n+1} -\alpha_{T,\mathrm{SR}} (T^{n+1}-T_{\mathrm{S}0}){\bm I}\bigr) \nonumber \\ 
    &+ \phi_\mathrm{I}(T) \mathbb{C}_\mathrm{IR}:\left(\bm\varepsilon^{n+1}-\bm\varepsilon^n - \alpha_{T,\mathrm{IR}} (T^{n+1}-T_\mathrm{fr}){\bm I} - \alpha_{\phi_\mathrm{I}} (S_\mathrm{I}(T^{n+1})-S_{\mathrm{I}0}){\bm I} \right),
\label{Int2}
\end{align}
with $\bm\varepsilon^{n+1}$ being now an unknown and $\bm\varepsilon^n$ being, as previously, the available input.

Our equation (\ref{Int2}) is term-by-term similar to the originally derived and presented equation (\ref{sigESI_3}). But in our case, the strain quantities $\bm\varepsilon^{n+1}$ and $\bm\varepsilon^n$ seem to have specified the meaning of the related strains $\bm\varepsilon_\mathrm{S}$ and $\bm\varepsilon_{\mathrm{S}0}$ from (\ref{sigESI_3}). 

An important notice for the component $\bm\varepsilon^n$ is still to be made. Thus, according to the above description of $\bm\varepsilon_{\mathrm{S}0}$, the input $\bm\varepsilon^n$ must be viewed as a strain of only a solid matrix, accrued prior to freezing. This stipulates the following {\em updating} scheme within the iterative solution process: Let the time step $n+1$ be the current one, let $\bm\varepsilon^n$ be the input and $\bm\varepsilon^{n+1}$ (and $T^{n+1}$) be the calculated solution. The appropriate update for the input $\bm\varepsilon^n$ to be used in the next step reads
\begin{equation}
\bm\varepsilon^n\leftarrow\bm\varepsilon^n+\bigl(1-S_\mathrm{I}(T^{n+1})\bigr)\bigl(\bm\varepsilon^{n+1}-\bm\varepsilon^n\bigr).
\label{update}
\end{equation}
Due to the factor $1-S_\mathrm{I}(T^{n+1})$, it is ensured that only the points ${\bm x}\in\Omega$ are updated which belong to the not-yet-frozen solid matrix, as required by \cite[cf. \autoref{subsec:BenchmarkM+freezing}]{Bluhm2011}.

%%%%%%%%%%%%%%%%%%%%%%%%%%%%%%%%%%%%%%%%%%%%%%%%%%%%%%%%%%%%%%%%%%%%%%%%%%%%%%%%%%%%%%%%%%%%%%%%%%%%%%%%%%%%%%%%%%%%
\section{Benchmarks for the model and the OGS code verification}
\label{sec:Benchmarks}
This section comprises six different benchmark problems with different levels of complexity. Note that they all refer to pure cases, i.\,e. the thermal problem, the hydraulical problem and the mechanical problem. However, in each of them the freezing feature introduces a nonlinear thermal (coupling) effect by itself.

%%%%%%%%%%%%%%%%%%%%%%%%%%%%%
\subsection{Benchmark tests for the T+freezing problem}
\label{subsec:BenchmarkT+freezing}
In this section, we consider the initial boundary value problem (IBVP) for the T+freezing equation (\ref{BalEn}), in which under zero deformation $(T)^\prime_\mathrm{S}$ turns into $\frac{\partial T}{\partial t}$ and where we also exclude the liquid flow (the term $\widetilde{\bm w}_\mathrm{LS}$):
\begin{equation}
    \left(
    (\varrho c_p)^\mathrm{eff}-\ell\varrho_\mathrm{IR}\frac{\mathrm{d}\phi_\mathrm{I}}{\mathrm{d}T}
    \right)\frac{\partial T}{\partial t}
    -\lambda^\mathrm{eff} \Delta T
    -(\lambda_\mathrm{IR}-\lambda_\mathrm{LR})\frac{\mathrm{d}\phi_\mathrm{I}}{\mathrm{d}T}|\nabla T|^2=Q_T({\bm x},t) \quad\mbox{in}\;\Omega,
\label{BalEnExt}
\end{equation}
\begin{equation}
\left\{
    \begin{tabular}{cll}
    $T=T_0({\bm x})$ & in $\Omega$ & (IC),   \\  [0.2cm]
    $T=T_1(t)$ & on $\Gamma_D$ & (Dirichlet BC),    \\  [0.2cm]
    $\frac{\partial T}{\partial{\bm n}}=G(t)$ & on $\Gamma_N$ & (Neumann BC),
\end{tabular}
\right.
\label{BalEnIBCs}
\end{equation}
for which three benchmark tests are developed and computed using OGS.\footnote{Note that Eq. (\ref{BalEnExt}) is nothing but (\ref{BalEn}), with the divergence term being expanded and the advective term neglected.} The developed code is verified as well. Verification scenarios differ by (a) the dimension of the considered problems ($\Omega\subset\mathbb{R}^n$, $n=1,2$ and $3$), (b) the type of the process modeled (melting, freezing) and, finally, (c) the kind of a reference solution (analytical, manufactured, numerical) used for comparison purposes. Thus,
\begin{itemize}
    \item In Section~\ref{T+freezing_Stefan}, we first consider and detail the so-called two-phase Stefan problem which models melting of a semi-infinite solid slab (in our case, an ice slab), see \cite{AlexSol}, and for which the closed-from analytical solution in $x\in(0,\infty)$ is available. Then, we apply (\ref{BalEnExt})--(\ref{BalEnIBCs}) to model such a melting process and solve the problem with OGS. This is done in the finite interval $x\in(0,4\,\mathrm{m})$ extracting the Dirichlet boundary conditions and the initial condition from the reference analytical data. The results obtained in $x\in(0,4\,\mathrm{m})$ at various time-steps for the two modeling approaches are compared.
    \item In Section~\ref{T+freezing_ManSol}, we verify the OGS code developed for solving (\ref{BalEnExt})--(\ref{BalEnIBCs}) using the notion of manufactured solution. Thus, we prescribe the temperature evolution which mimics a combined melting-freezing process in a porous medium. The domain of interest is taken to be a unit square. The given temperature function is plugged in (\ref{BalEnExt})--(\ref{BalEnIBCs}) to recover the corresponding right-hand side data, such as the source term $Q_T$, as well as the initial temperature $T_0$ and the function $T_1$ for the Dirichlet boundary condition. The recovered data set is then used in the OGS code for solving (\ref{BalEnExt})--(\ref{BalEnIBCs}) as input, and the obtained numerical solution is compared with the prescribed (manufactured) counterpart.
    \item In Section \ref{T+freezing_SoilFr}, we use (\ref{BalEnExt})--(\ref{BalEnIBCs}) to model ice forming in a cylindrical soil specimen around a borehole heat exchanger (BHE) which contains a refrigerant of sub-zero temperature. This temperature is used to prescribe a Dirichlet boundary condition on the specimen boundary adjacent to the BHE, what triggers cooling and consequent freezing of soil whose initial temperature is positive. Simulations are performed using both the OGS and the FreeFem++ open source finite element package \cite{FreeFem} and the results are compared, thus enabling also verification of the developed codes.
\end{itemize}
For the sake of simplicity but without loss of generality, in all considered examples, we opt for dealing in Eq. (\ref{BalEnExt})--(\ref{BalEnIBCs}) only with the Dirichlet boundary conditions.

%%%%%%%%%%%%%%%
\subsubsection{Two-phase Stefan problem (melting of a semi-infinite slab)}
\label{T+freezing_Stefan}
\begin{figure}[h!]
\begin{center}
\includegraphics[width=0.85\textwidth]{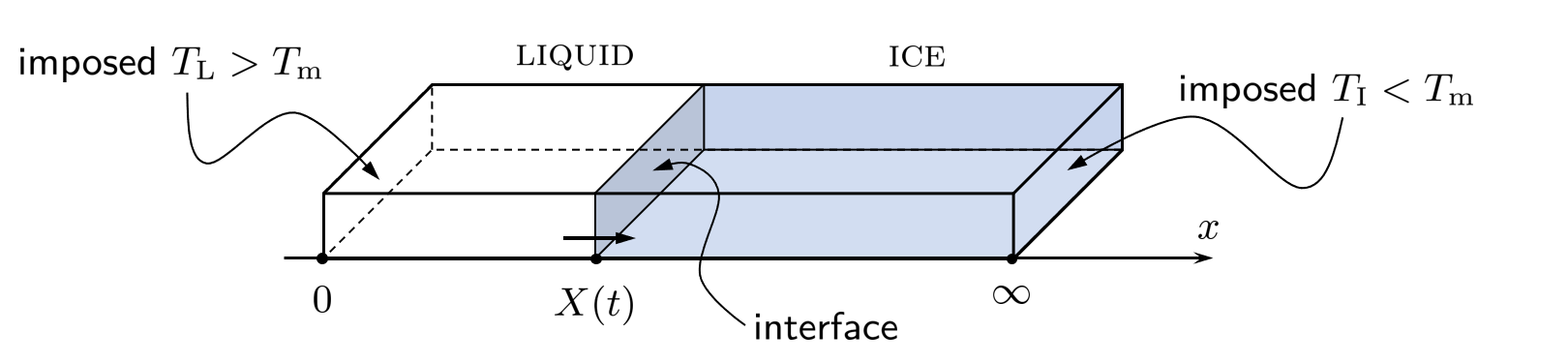}
\end{center}
\caption{Sketch of a semi-infinite melting slab as a physical situation modelled by the two-phase Stefan problem.}
\label{fig_0}
\end{figure}

Physically, the two-phase Stefan problem models a semi-infinite slab, $0\leq x<\infty$, initially solid at sub-zero temperature $T_\mathrm{I}<T_\mathrm{fr}$, which starts melting by imposing a constant positive temperature $T_\mathrm{L}>T_\mathrm{fr}$ at $x=0$. We expect that during melting, an interface between the two phases given by $x=X(t)$ such that $X(0)=0$ advances to the right. The situation is sketched in \autoref{fig_0}, where the boundary condition for $T$ at infinity, that is, $\lim_{x\rightarrow\infty}T(x,t)=T_\mathrm{I}$ is also depicted.\footnote{Note that the homogeneous Neumann boundary conditions may also be physically appropriate, but the exact analytical solution cannot be derived in this case.} Finally, it is assumed that the corresponding phases are characterized by the material parameters $\varrho_\mathrm{LR},\lambda_\mathrm{LR},c_{p\mathrm{L}}$ and $\varrho_\mathrm{IR},\lambda_\mathrm{IR},c_{p\mathrm{I}}$.

We skip the detailed mathematical formulation of the problem, only referring to the solution equations.\footnote{A comprehensive treatment is available in \cite{AlexSol}.} Thus, the temperature evolution during the process is described as follows:
\vspace{-3mm}
\begin{equation}
T(x,t):=
\left\{
\begin{tabular}{lr}
    $T_\mathrm{L}-(T_\mathrm{L}-T_\mathrm{fr})\displaystyle
    \frac{\mathrm{erf}
    \left(\frac{x}{2\sqrt{a_\mathrm{L}t}}\right)}
    {\mathrm{erf}(\Lambda)}, \quad 0<x\leq X(t),\;t>0$ & (\textsc{Liquid}),   \\  [0.2cm]
    $T_\mathrm{I}+(T_\mathrm{fr}-T_\mathrm{I})\displaystyle
    \frac{\mathrm{erfc}\left(\frac{x}{2\sqrt{a_\mathrm{I}t}}
    -(1-\varrho^\star)a^\star\Lambda\right)}
    {\mathrm{erfc}(\varrho^\star a^\star\Lambda)}, \quad X(t)\leq x<\infty,\;t>0$ & (\textsc{Ice}),
\end{tabular}
\right.
\label{StT}
\end{equation}
where $a_\mathrm{L}:=\frac{\lambda_\mathrm{LR}}{\varrho_\mathrm{LR}c_{p\mathrm{L}}}$ and $a_\mathrm{I}:=\frac{\lambda_\mathrm{IR}}{\varrho_\mathrm{IR}c_{p\mathrm{I}}}$ are the thermal diffusivity of the corresponding phases, $\varrho^\star:=\frac{\varrho_\mathrm{LR}}{\varrho_\mathrm{IR}}$ and $a^\star:=\sqrt{\frac{a_\mathrm{L}}{a_\mathrm{I}}}$ are dimensionless parameters, $X(t):=2\Lambda\sqrt{a_\mathrm{L}t}$, $t>0$ is the position of a melt front (an interface), and $\Lambda>0$ is the root of the transcendental equation
\vspace{-2mm}
\begin{equation}
    \frac{\mathrm{St}_\mathrm{L}}{\Lambda\exp(\Lambda)^2\mathrm{erf}(\Lambda)}
    -\frac{St_\mathrm{I}}{\varrho^\star a^\star\Lambda\exp(\varrho^\star a^\star\Lambda)^2\mathrm{erfc}(\varrho^\star a^\star\Lambda)}
    =\sqrt{\pi},
\label{StLamb}
\end{equation}
where, in turn, $St_\mathrm{L}>0$ and $St_\mathrm{I}>0$ are the so-called Stefan numbers defined as
\begin{equation}
    St_\mathrm{L}:=\frac{c_{p\mathrm{L}}(T_\mathrm{L}-T_\mathrm{fr})}{\ell},
    \quad
    St_\mathrm{I}:=\frac{c_{p\mathrm{I}}(T_\mathrm{fr}-T_\mathrm{I})}{\ell},
\label{StNum}
\end{equation}
with, finally, $\ell$ being the latent heat of fusion. 

\begin{figure}[h!]
\begin{center}
\includegraphics[width=1.0\textwidth]{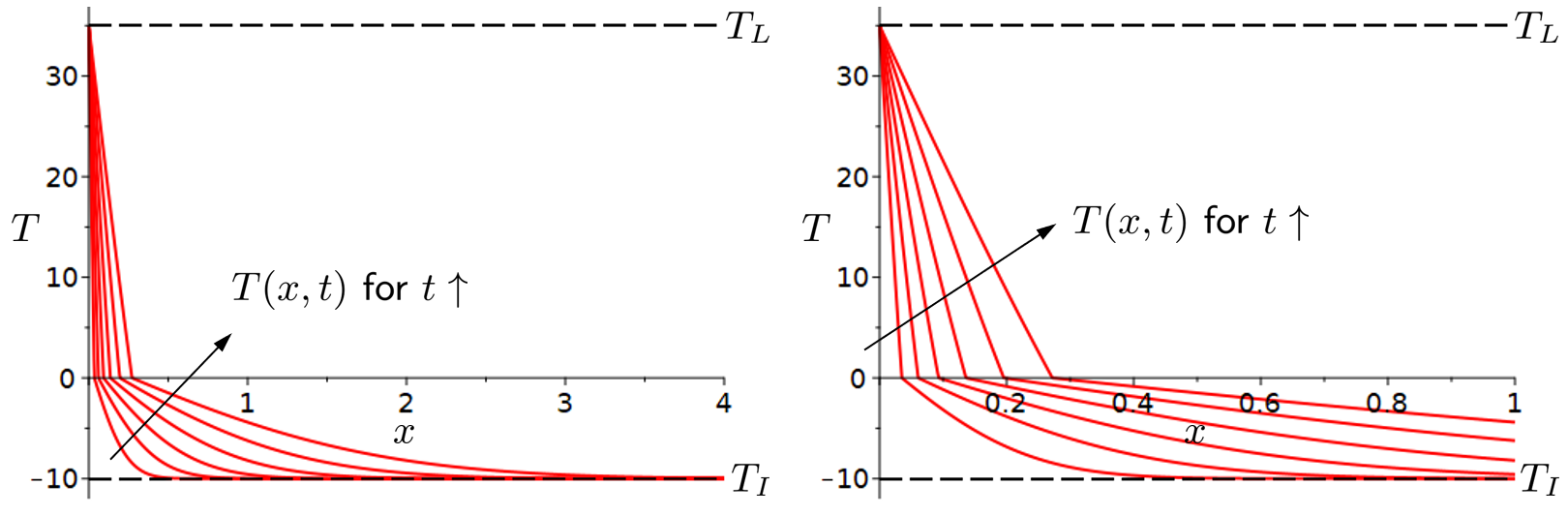}
\end{center}
\caption{Solution (\ref{StT}) at time-steps $t\in\{4\,\mathrm{h},12\,\mathrm{h},28\,\mathrm{h},60\,\mathrm{h},124\,\mathrm{h},$ $240\,\mathrm{h}\}$ in the intervals $x\in[0,4\,\mathrm{m}]$ (on the left) and $x\in[0,1\,\mathrm{m}]$ (on the right); dashed lines depict the prescribed $T_\mathrm{L}$ and $T_\mathrm{I}$ in $^\circ$C.}
\label{fig_1}
\end{figure}

To calculate all the induced quantities in (\ref{StT})--(\ref{StNum}), we use the material parameters presented in Table~\ref{table_MatData_Stefan}. In this case, $\Lambda\approx 0.3933292421$ and the corresponding temperature evolution within the melting process of a slab given by equations~(\ref{StT}) can be visualized, see \autoref{fig_1}. In the plots, $T$ is depicted at different time-steps $t\in\{4\,\mathrm{h},12\,\mathrm{h},28\,\mathrm{h},60\,\mathrm{h},124\,\mathrm{h},$ $240\,\mathrm{h}\}$ and in the finite intervals $x\in[0,4\,\mathrm{m}]$ (the left plot) and $x\in[0,1\,\mathrm{m}]$ (the right plot). The former case intends to illustrate that the boundary condition at infinity, $\lim_{x\rightarrow\infty}T(x,t)=T_\mathrm{I}$ is indeed fulfilled, whereas the latter one is only the zoom into the region in which the comparison between the analytical and the numerical results will later be performed.

\begin{table}[h]
\small
\centering
\begin{tabular}{ l | l }
	Liquid phase & Ice phase \\
	\hline
	$\varrho_\mathrm{LR}=1000$ kg/m$^3$  &  $\varrho_\mathrm{IR}=920$ kg/m$^3$ \\
	$c_{p\mathrm{L}}=4190$ J/(kg K)  &  $c_{p\mathrm{I}}=2090$ J/(kg K) \\
	$\lambda_\mathrm{LR}=0.58$ W/(m K)  &  $\lambda_\mathrm{IR}=2.2$ W/(m K) \\
	& $\ell=3.34\cdot 10^5$ J/kg \\
	\hline
	$T_\text{L}=35$ $^\circ$C (308.15 K)
	& $T_\text{I}=-10$ $^\circ$C (263.15 K) \\
	\hline
	\multicolumn{2}{l}{Freezing temperature, $T_\mathrm{fr}=0$ $^\circ$C (273.15 K)}\\
	\hline
\end{tabular}
\caption{Material properties and parameters for the Stefan problem.}
\label{table_MatData_Stefan}
\vspace{-10mm}
\end{table}

We now model the slab melting process using our IBVP for the T+freezing equation (\ref{BalEnExt}) implemented in the OGS code and compare the simulated results with the analytical ones presented in \autoref{fig_1}. In the computations for (\ref{BalEnExt}), we restrict ourselves to the interval $x\in[0,4\,\mathrm{m}]$. The boundary condition for the unknown $T^h$ at $x=0$ is $T_\mathrm{L}$. At $x=4\,\mathrm{m}$, we set $T^h\coloneqq T(4\,\mathrm{m},t)$, where $T(4\,\mathrm{m},t)$ is the ice-phase part of the Stefan solution in (\ref{StT}) evaluated at $x=4$. Imposing the initial condition is somewhat more involving: The direct use of the original IC $T^h\coloneqq T_\mathrm{I}$ at $t=0,x\geq0$ -- given its incompatibility with the BC $T_\mathrm{L}$ at $x=0,t\geq0$, and the fact that we have a problem with a (strong) boundary layer -- will result in convergence issues at the first time step of computation. Therefore, in our solution process, we opt for taking $t_0:=3600\,\mathrm{s}$ as the initial moment and the initial condition for $T^h$ is hence taken to be the Stefan solution $T(x,t_0)$. The simulation time interval is $t\in[3600\,\mathrm{s},864000\,\mathrm{s}]=[1\,\mathrm{h},240\,\mathrm{h}]$, and the time increment $\Delta t:=36\,\mathrm{s}$. The spatial discretisation of $x\in[0,4\,\mathrm{m}]$ is uniform and uses $\Delta x=0.005\,\mathrm{m}$. Finally, we set the porosity $\phi=1$ and compute two cases in terms of the Sigmoid function coefficient $k$, namely $k\coloneqq2$ (Case 1) and $k\coloneqq5$ (Case 2).

\begin{figure}[h!]
\begin{center}
\includegraphics[width=1.0\textwidth]{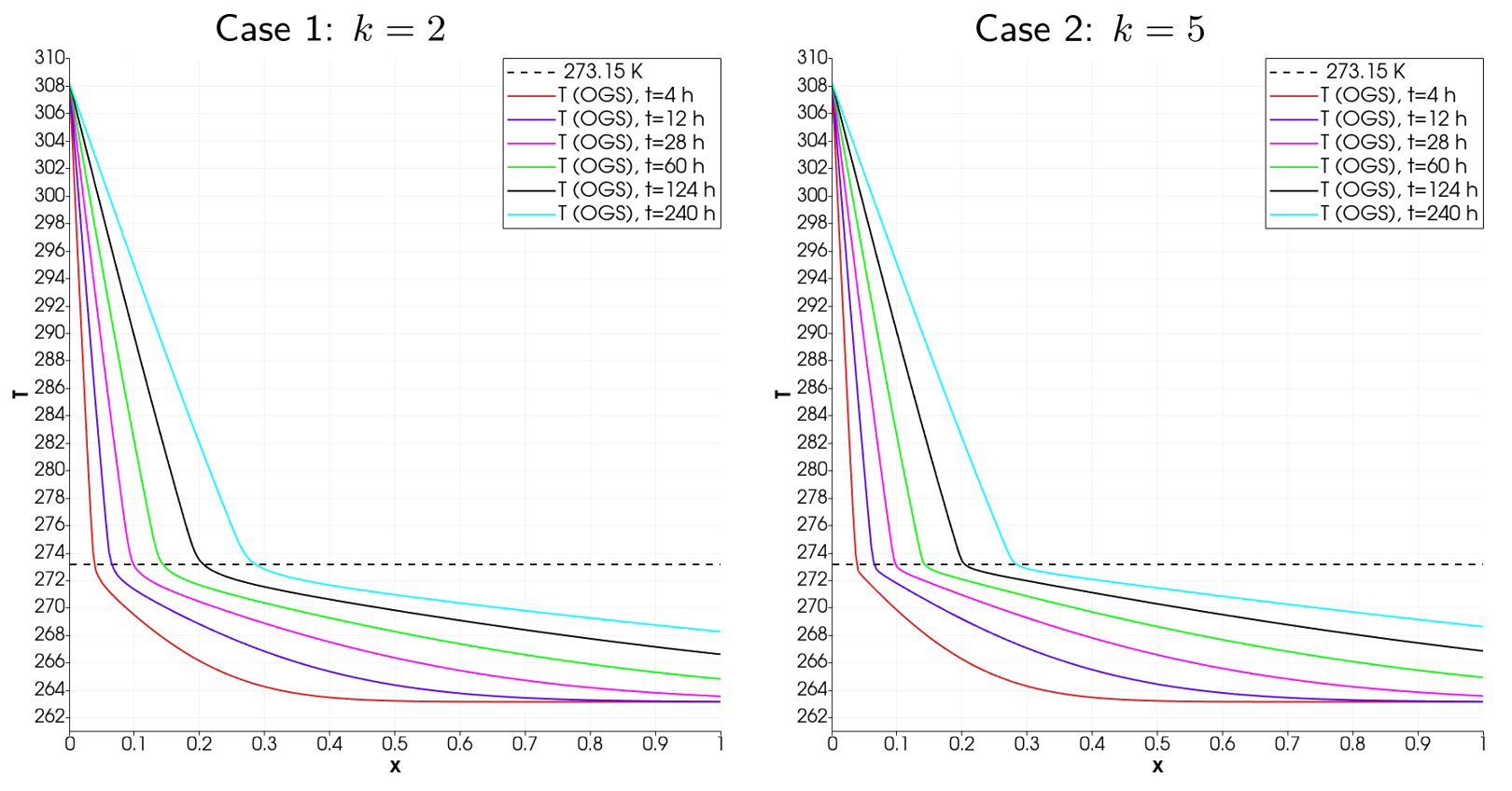}
\end{center}
\caption{The OGS solution of the IBVP for the T+freezing equation~(\ref{BalEnExt}) at time-steps $t\in\{4\,\mathrm{h},12\,\mathrm{h},28\,\mathrm{h},$ $60\,\mathrm{h},124\,\mathrm{h},240\,\mathrm{h}\}$ in the interval $x\in[0,1\,\mathrm{m}]$ for the two cases of $k$; temperature is given in the unit Kelvin.}
\label{fig_1a}
\end{figure}

The results obtained with OGS for both cases at time steps $t\in\{4\,\mathrm{h},12\,\mathrm{h},28\,\mathrm{h},60\,\mathrm{h},$ $124\,\mathrm{h},240\,\mathrm{h}\}$ are depicted in \autoref{fig_1a}, where we restrict the solution plots to the interval $[0,1\,\mathrm{m}]$. The freezing temperature $T_\mathrm{fr}$ is also plotted to make the interface visible. It can be seen that the general trend of temperature evolution in the slab in either simulated case is the same as the reference one in \autoref{fig_1}, right. For the larger $k$ (Case 2), the temperature kink at the interface becomes sharper, as expected, since $k$ governs the steepness of $S_\mathrm{I}$ in the transition zone and thus, the induced spatial extension of this zone.

\begin{figure}[h!]
\begin{center}
\includegraphics[width=1.0\textwidth]{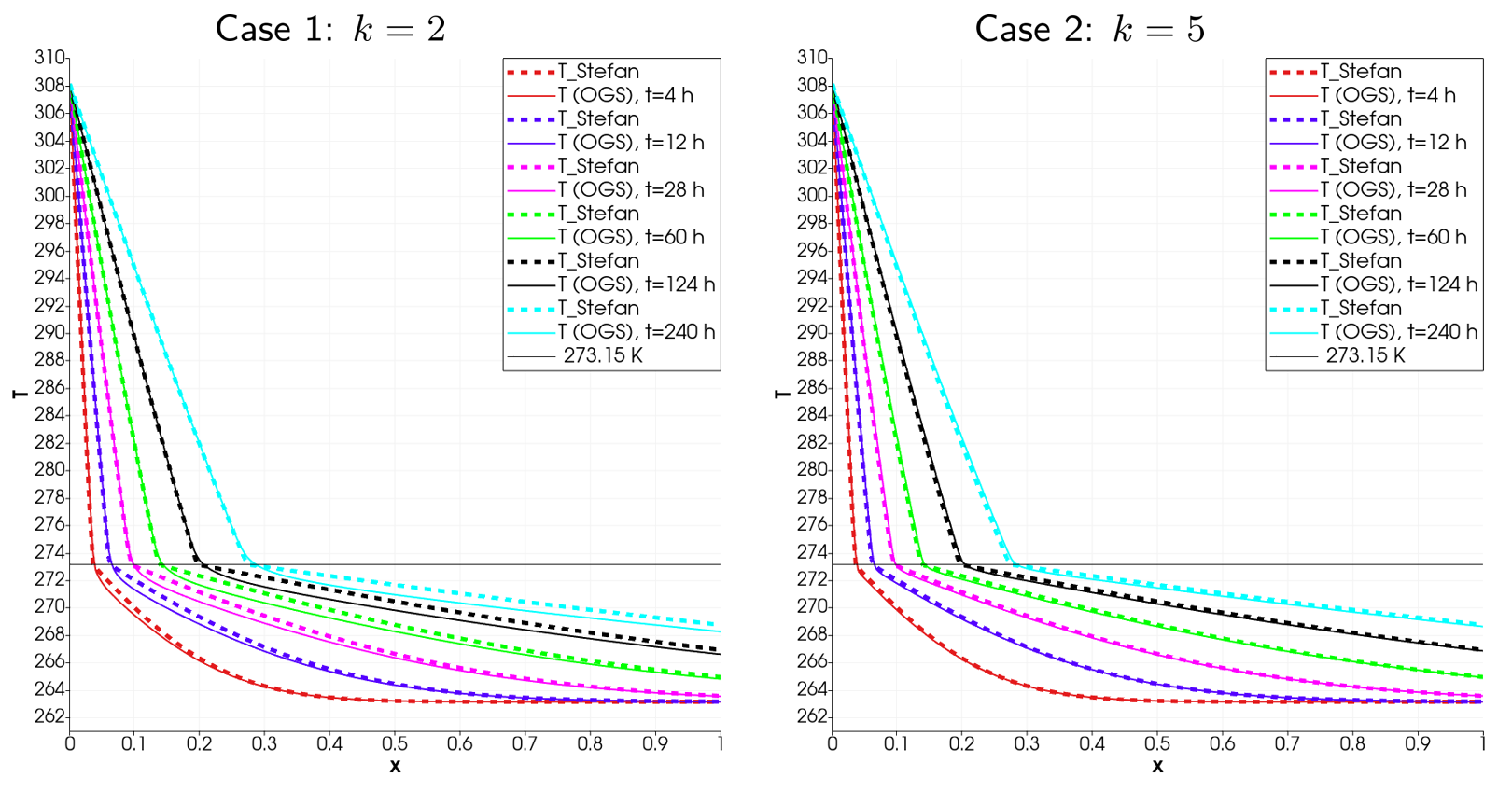}
\end{center}
\caption{Comparison of the analytical solution (\ref{StT}) of two-phase Stefan problem to the numerical OGS solution of the IBVP for the T+freezing equation (\ref{BalEnExt}), obtained in two cases of $k$; the temperature is given in the unit Kelvin.}
\label{fig_1b}
\end{figure}

\autoref{fig_1b} presents a comparison of the analytical solution~(\ref{StT}) of Stefan problem and the OGS solution of the IBVP for the T+freezing equation (\ref{BalEnExt}) obtained in two cases of $k$ in the interval $x\in[0,4\,\mathrm{m}]$. Again, both representations are restricted to $x\in[0,1\,\mathrm{m}]$. 
Note that we have applied two mathematical models to the very same physical problem of slab melting -- the original two-phase Stefan problem and the one given by the T+freezing equation. Hence, neither can be taken as a reference. Despite this fact we observe that the quantitative discrepancy of the results is minor, especially in Case 2, with the increased magnitude of $k$.

%%%%%%%%%%%%%%%
\subsubsection{2D problem with manufactured solution}
\label{T+freezing_ManSol}
When the analytical solution of the formulation of interest is not available (this is opposite to the situation in paragraph \ref{T+freezing_Stefan}), the concept of a manufactured solution can also be used for code verification purposes. 

To illustrate the idea, consider the boundary value problem for the Poisson equation in two dimensions:
\begin{equation}
\left\{
\begin{tabular}{cl}
$-\Delta u(x,y)=f(x,y)$ & in $\Omega$,   \\  [0.2cm]
$u=g$ & on $\Gamma_D$, \\
$\frac{\partial u}{\partial{\bm n}}=h$ & on $\Gamma_N$,
\end{tabular}
\right.
\label{Poisson}
\end{equation}
where $\Delta=\frac{\partial^2}{\partial x^2}+\frac{\partial^2}{\partial y^2}$ is the Laplace operator, $\Omega$ is an open and bounded domain, $\Gamma_D$ and $\Gamma_N$ are non-overlapping parts of the domain's boundary $\partial\Omega$ where the Dirichlet and Neumann boundary conditions are formulated, respectively. The {\em direct} solution task reads: for the given data $f,g,h$, find the unknown $u$ that satisfies (\ref{Poisson}). 

Assume the inverse situation: we take an arbitrary\footnote{and such that it is twice continuously differentiable in $\Omega\cup\partial\Omega$} $u$ in $\Omega$ and plug it in Eq. (\ref{Poisson}), thus recovering the corresponding right-hand sides $f$ in $\Omega$ and $g,h$ on $\partial\Omega$: 
\begin{equation*}
f(u):=-\Delta u, \quad g(u):=u|_{\Gamma_D} \quad \mathrm{and} \quad h(u):=\tfrac{\partial u}{\partial{\bm n}}|_{\Gamma_N}.
\end{equation*}
Note that $a|_{\Gamma_b}$ means the restriction of $a$ to $\Gamma_b$. Such $u$ is called a manufactured solution. If we were now to solve (\ref{Poisson}) with the $u$-induced right-hand sides $f,g,h$ e.\,g.\ numerically, it would be straightforward to compare the computed discretization $u^h$ -- both qualitatively and quantitatively -- with the already available $u$, thus verifying the corresponding ingredients of the numerical implementation (a code, an algorithm, a method etc.). Importantly, well-posedness of problem (\ref{Poisson}) with the given $u$-induced data $f$, $g$ and $h$ is guaranteed, since compatibility between these functions is automatically assured. 

In the following, we apply the described approach to problem (\ref{BalEnExt})--(\ref{BalEnIBCs}). In doing so, we restrict ourselves to a two-dimensional formulation in the unit square $\Omega:=(0,L)\times(0,L)$ with $L=1\,\mathrm{m}$. We also opt for dealing only with the Dirichlet type boundary data, that is, $\Gamma_D:=\partial\Omega$ is assumed. Physically, the simultaneous ice melting-forming process is envisioned, which is to be mimicked by the choice of the manufactured $T=T(x,y,t)$, $t\in(0,1]\,\mathrm{s}$. We will provide the expressions for $Q_T$, $T_0$ and $T_1$ required in (\ref{BalEnExt})--(\ref{BalEnIBCs}) explicitly. Our OGS implementation is tested when $\Omega$ is discretized by linear triangles and bi-linear quadrilaterals such that the corresponding FE solution $T^h$ at any fixed time step is in $P^1$ and $Q^1$ spaces, respectively. The result presented are for the case of $Q^1$-based discretization $T^h$ only. Finally, the material data and parameters in (\ref{BalEnExt}) used in the numerical experiments are depicted in Table \ref{table_MatData_Tfreezing}.

\begin{table}[h]
\small
\centering
\begin{tabular}{ l | l | l  }
	solid phase & liquid phase & ice phase \\
	\hline
	$\varrho_\mathrm{SR}=2000$ kg/m$^3$ &  $\varrho_\mathrm{LR}=1000$ kg/m$^3$  &  $\varrho_\mathrm{IR}=920$ kg/m$^3$ \\
	$c_{p\mathrm{S}}=900$ J/(kg K) &  $c_{p\mathrm{L}}=4190$ J/(kg K)  &  $c_{p\mathrm{I}}=2090$ J/(kg K) \\
	$\lambda_\mathrm{SR}=1.1$ W/(m K) &  $\lambda_\mathrm{LR}=0.58$ W/(m K)  &  $\lambda_\mathrm{IR}=2.2$ W/(m K) \\
	 & & $\ell=3.34\cdot 10^5$ J/kg \\
	\hline
	\multicolumn{3}{l}{Porosity, $\phi=0.5$}\\
	\multicolumn{3}{l}{Sigmoid function $S_\mathrm{I}$ coefficient, $k=2$}\\
	\multicolumn{3}{l}{Freezing temperature, $T_\mathrm{fr}=0$ $^\circ$C (273.15 K)}\\
	\hline
\end{tabular}
\caption{Material properties and parameters in the manufactured solution benchmark.}
\label{table_MatData_Tfreezing}
\end{table}

\begin{figure}[h!]
\begin{center}
\includegraphics[width=1.0\textwidth]{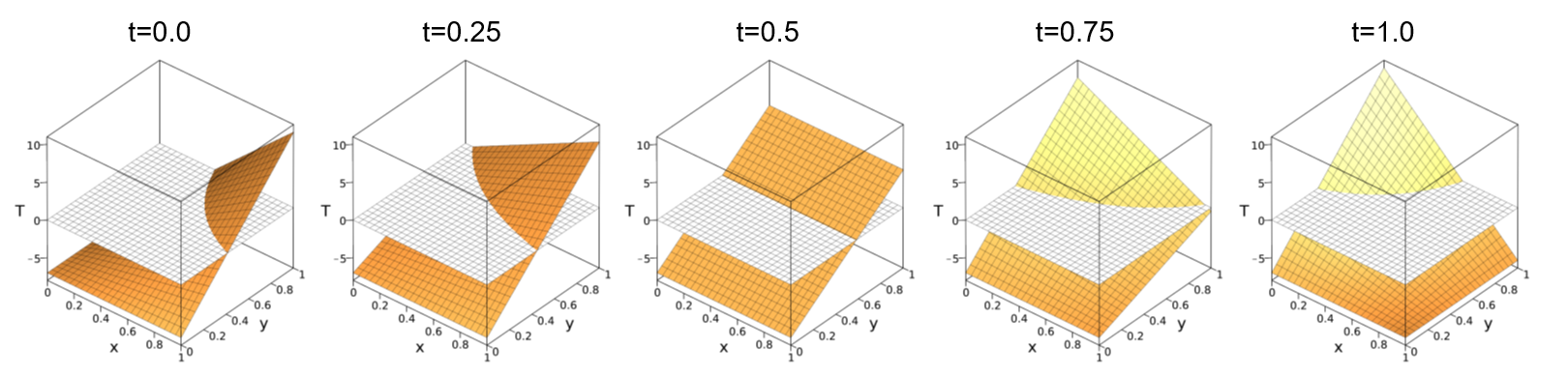}
\end{center}
\caption{Snapshots of $T$ given by (\ref{ManufTcase3}) in degrees Celsius. The plane in white color represents zero (freezing) temperature $T_\mathrm{fr}=0$ $^\circ$C; the zero-level set of $T$ mimics the interface between ice and water fractions, which moves in time.}
\label{fig_5}
\end{figure}

The manufactured solution $T$ for (\ref{BalEnExt})--(\ref{BalEnIBCs}) in this case is given by
\begin{equation}
    T(x,y,t):=b\left[xy\cos\left(\omega t\right) 
    +(L-x)y\sin\left(\omega t\right)\right]+c,\quad \omega:=\tfrac{1}{2}\tfrac{\pi}{\mathrm{s}}, 
\label{ManufTcase3}
\end{equation}
where $b=17\,\tfrac{\mathrm{K}}{\mathrm{m}^2}$ and $c=(-7+273.15)$ K. At every time step, it is a simple  bi-linear function of $x$ and $y$. The snapshots of $T$ are depicted in \autoref{fig_5}, where the horizontal plane stands for the melting temperature, such that the physical meaning of $T$ defined by (\ref{ManufTcase3}) can be easily grasped: the evolution of temperature mimics both ice formation and ice melting within the domain, which is initially partly occupied by ice and liquid. Its zero level set (intersection of the plane and $T$) represents the interface between the two fractions.

Using (\ref{ManufTcase3}), the right-hand side data for (\ref{BalEnExt})--(\ref{BalEnIBCs}) is recovered. The source term $Q_T$ in $\Omega$ reads:
\begin{gather}
    Q_T:=\left(
    (\varrho c_p)^\mathrm{eff}-\ell\varrho_\mathrm{IR}\frac{\mathrm{d}\phi_\mathrm{I}}{\mathrm{d}T}
    \right)
    \omega by
    \left[
    (L-x)\cos\left(\omega t\right) 
    -x\sin\left(\omega t\right)
    \right] \notag
    \\
    -(\lambda_\mathrm{IR}-\lambda_\mathrm{LR})\frac{\mathrm{d}\phi_\mathrm{I}}{\mathrm{d}T}b^2
    \left[
    y^2(1-\sin(2\omega t))+x(L-x)\sin(2\omega t)+L\left(x-\tfrac{L}{2}\right)\cos(2\omega t) \right. \nonumber \\
    \left.+x^2-L x+\tfrac{L^2}{2}\right].
\label{ManufQTcase3}
\end{gather}
The initial condition function $T_0$ in $\Omega$ is as follows: 
\begin{equation*}
    T_0(x,y):=T(x,y,0)
    =bxy+c.
\end{equation*}
Finally, assuming $\Gamma_D$ to be composed of all four sides of the unit square, the boundary condition function $T_1$ is also obtained:
\begin{equation*}
T_1(t):=
\left\{
    \begin{tabular}{ll}
    $c$, & on $\{x\in(0,L),y=0\}$,   \\  [0.2cm]
    $by\cos\left(\frac{1}{2}\pi t\right)+c$, & on $\{x=L,y\in(0,L)\}$, \\  [0.2cm]
    $b\left[x\cos\left(\frac{1}{2}\pi t\right)
    +(1-x)\sin\left(\frac{1}{2}\pi t\right)\right]+c$, & on $\{x\in(0,L),y=L\}$, \\  [0.2cm]
    $by\sin\left(\frac{1}{2}\pi t\right)+c$, & on $\{x=0,y\in(0,L)\}$.
\end{tabular}
\right.
\end{equation*}

\begin{figure}[h!]
\begin{center}
\includegraphics[width=1.0\textwidth]{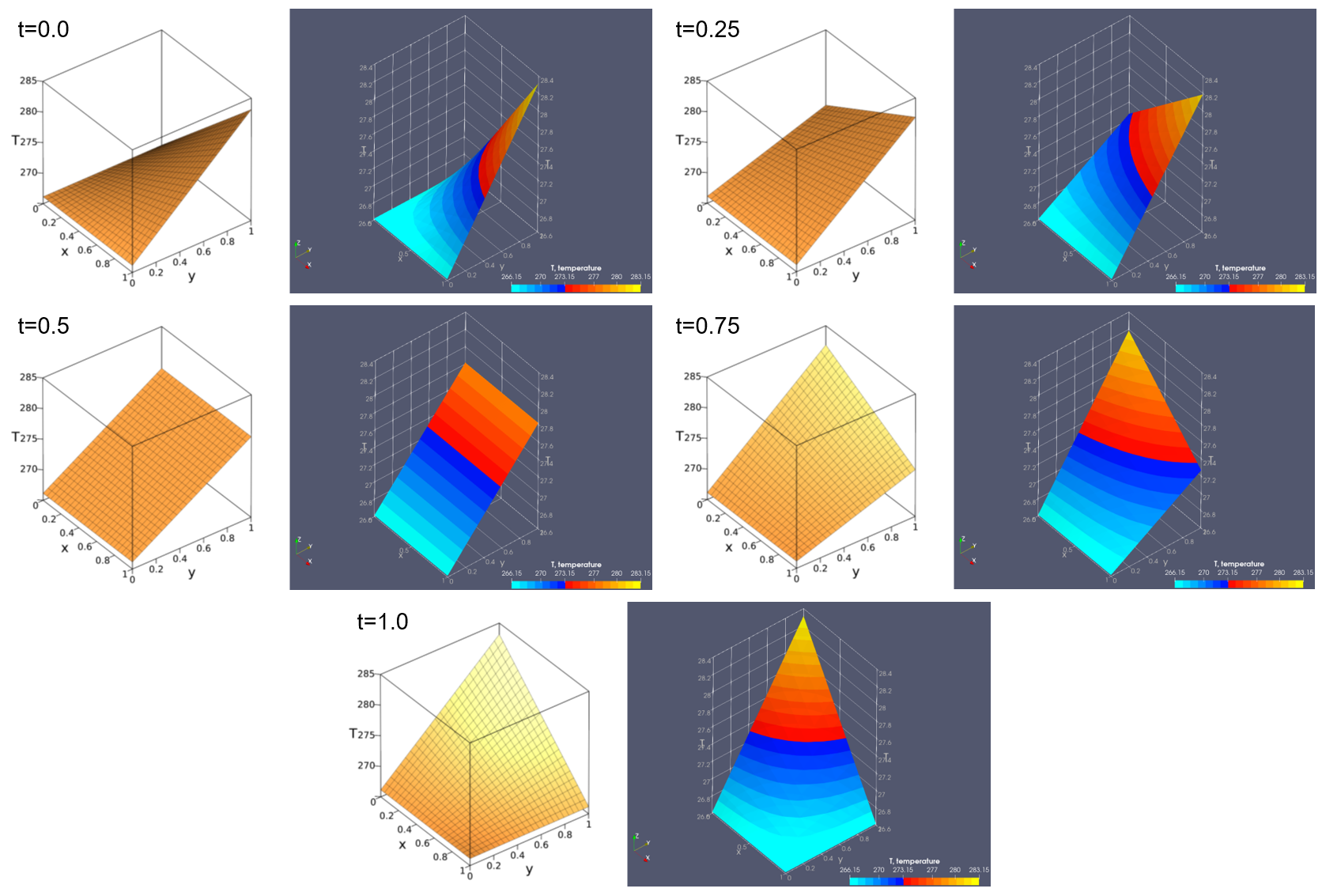}
\end{center}
\caption{Comparison of the manufactured solution (left) with the numerical one (right) computed by OGS at different time steps; temperature is given in the unit Kelvin.}
\label{fig_5a}
\end{figure}

\autoref{fig_5a} depicts a comparison of the snapshots of $T$ given by Eq. (\ref{ManufTcase3}) and the solution $T^h$ to (\ref{BalEnExt})--(\ref{BalEnIBCs}) obtained with OGS. The temperature is given in the unit Kelvin. Note that we rescaled the vertical range of the OGS solutions in the ParaView plots by factor $10^{-1}$, to make the comparison feasible. Also, we tuned the color legend in the plots such that the ice and water fractions can be identified visibly. Already a qualitative similarity of the reference results and the numerical ones is evident.

%%%%%%%%%%%%%%%
\subsubsection{3D-axisymmetric problem of soil freezing around a BHE}\label{T+freezing_SoilFr}

\begin{figure}[h]
\begin{center}
\includegraphics[width=1.0\textwidth]{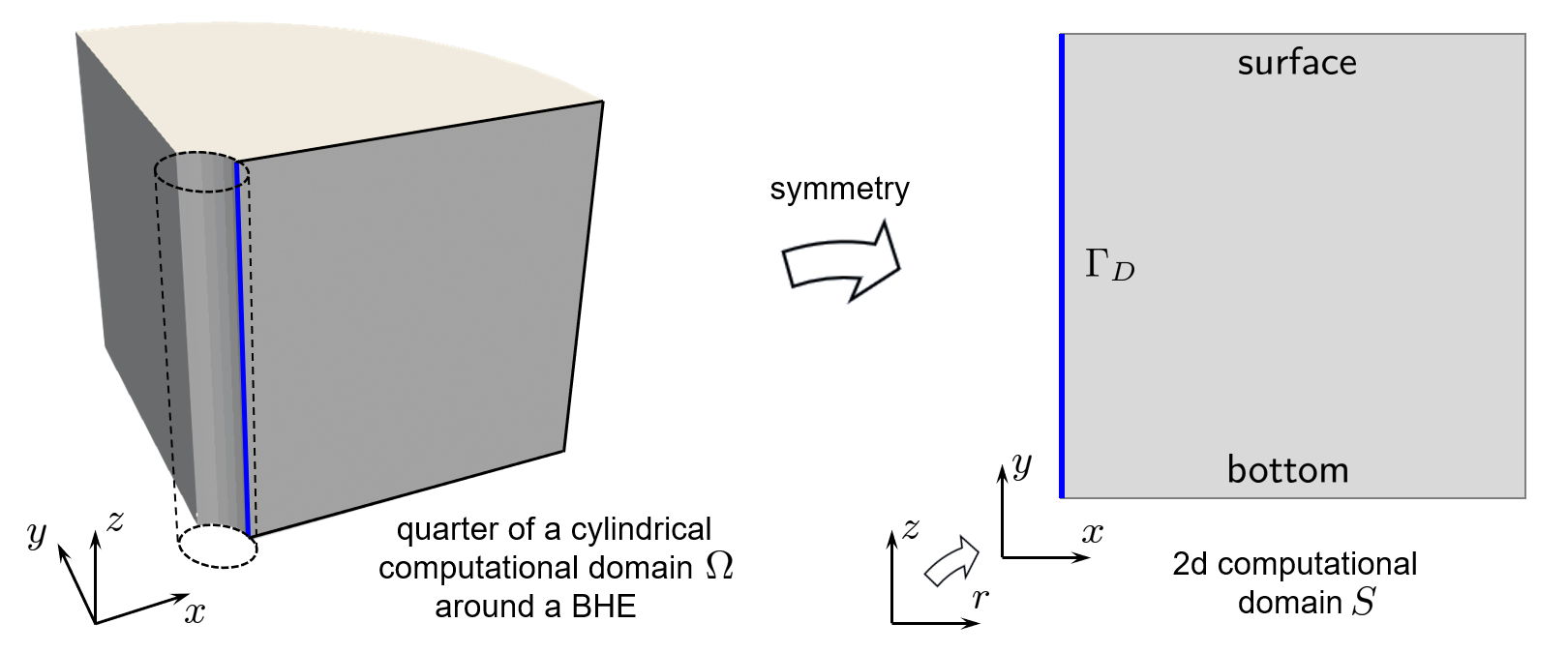}\\
\end{center}
\caption{On the left: quarter of a cylindrical soil block around a BHE (quarter of a 3-dimensional domain $\Omega$); on the right: reduction to a 2D problem with the corresponding computational domain $S$, where also $(r,z)\in S$ are re-denoted as $(x,y)$.}
\label{fig_8}
\end{figure}

In this section, using Equations (\ref{BalEnExt})--(\ref{BalEnIBCs}), we model a heat transfer process, focusing specifically on ice formation in a cylindrical soil specimen around a borehole heat exchanger (BHE), see the left plot in \autoref{fig_8}. It contains a refrigerant of sub-zero temperature. This temperature is used to prescribe a Dirichlet boundary condition on the specimen boundary adjacent to the BHE. That should trigger cooling and consequent freezing of water-saturated soil whose initial temperature was positive. 

Simulations are performed using both our OGS platform and the FreeFem++ open source finite element code \cite{FreeFem}, thus enabling code verification for solving (\ref{BalEnExt})--(\ref{BalEnIBCs}). Due to the domain and problem symmetry, we reduce the three-dimensional formulation to a two-dimensional one, as also sketched in \autoref{fig_8}. In the context of the corresponding weak form, this implies passing from three dimensional integration in the Cartesian coordinate system first to the integration using the cylindrical coordinates, which is then reduced to two-dimensional integration:
\begin{equation*}
    \int_\Omega f(x,y,z)\mathrm{d}x\mathrm{d}y\mathrm{d}z
    =\int_\Omega\widehat{f}(r,\theta,z)r\mathrm{d}r\mathrm{d}\theta\mathrm{d}z
    =2\pi\int_S \widehat{f}(r,z)r\mathrm{d}r\mathrm{d}z,
\end{equation*}
where $\widehat{f}(r,\theta,z):=f(r\cos(\theta),r\sin(\theta),z)$, and we also assumed $\Omega=S\times 2\pi$. In the following, the variables $(r,z)\in[0.25\,\mathrm{m},16.25\,\mathrm{m}]\times[-16\,\mathrm{m},0\,\mathrm{m}]=:S$ are to be re-denoted as $(x,y)$ respectively, representing the length of $S$ in the radial direction and the depth in the vertical direction, see \autoref{fig_8}, right. Note these new $(x,y)$ coordinates are unrelated to the original 3D formulation and the coordinate notations.

In our computations, the material data from Table \ref{table_MatData_Tfreezing} is used. The initial condition for $T$ in $S$ is assumed to be a positive function which decays linearly from surface to bottom: 
\begin{equation*}
    T_0(x,y):=-\frac{T_\mathrm{surf}-T_\mathrm{bot.}}{H_\mathrm{bot}}y
    +T_\mathrm{surf},
\end{equation*}
where $T_\mathrm{surf}:=25$ $^\circ$C and $T_\mathrm{bot}=10$ $^\circ$C are the temperatures at the surface and at the depth of $H_\mathrm{bot}=-16$ m, respectively.

\begin{figure}%[h]
\begin{center}
\includegraphics[width=1.0\textwidth]{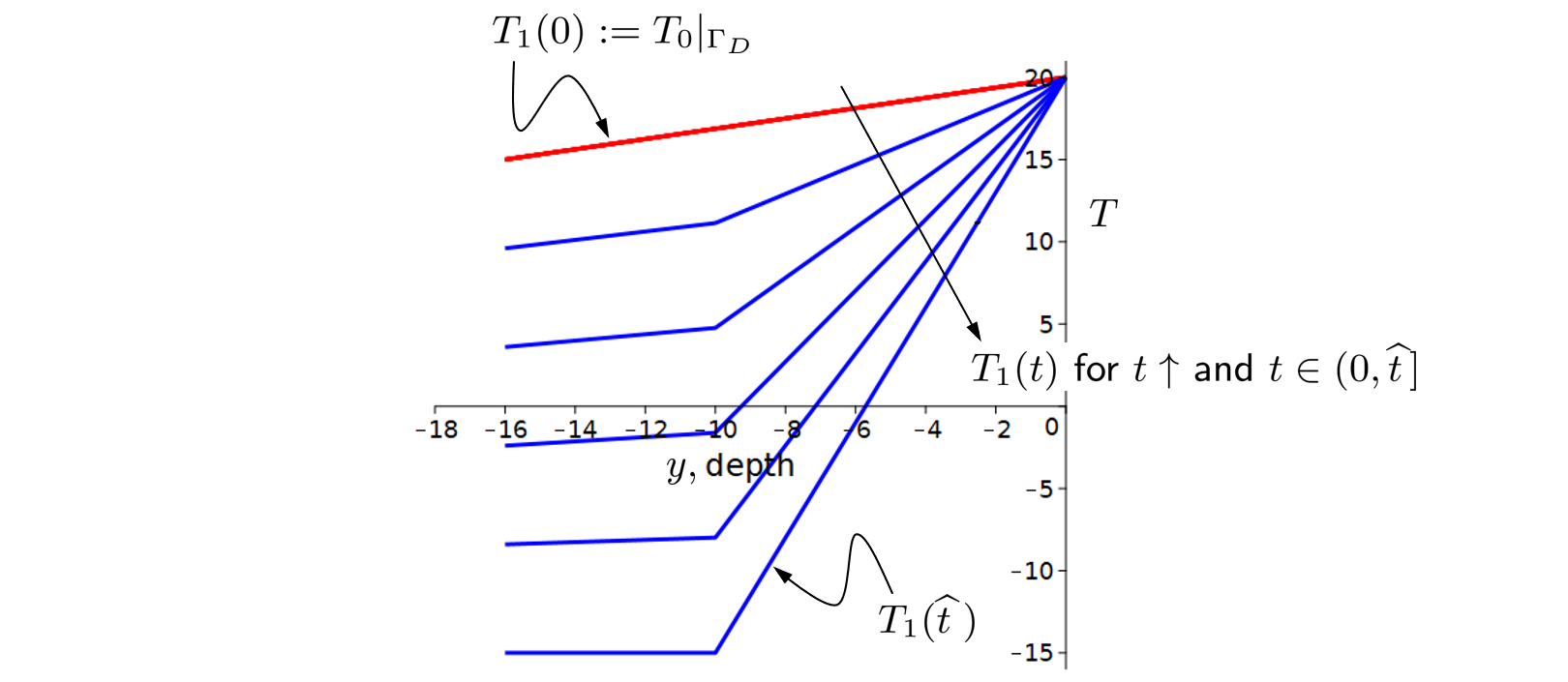}\\
\end{center}
\caption{Temperature evolution prescribed on $\Gamma_D$ within the time interval $t\in[0,\widehat{t}\,]$, where $\widehat{t}:=10$ hours; the horizontal and vertical tickmarks are in meters and $^\circ$C, respectively.}
\label{fig_9}
\end{figure}

For modeling the (time-dependent) boundary conditions on $\Gamma_D$ of $S$, we assume that within the first $\widehat{t}:=10$ hours, the temperature on $\Gamma_D$ drops continuously from $T_0|_{\Gamma_D}$ to the values prescribed by some continuous piece-wise linear function of $y$ and such that at the last depth segment $y\in[H_\mathrm{bot},H_\mathrm{fr}]$, where $H_\mathrm{fr}:=-10$ m, it becomes negative. (To recall, the latter mimics the impact of the BHE refrigerant with sub-zero temperature.) \autoref{fig_9} sketches the situation, whereas explicitly, for $t\in[0,\widehat{t}\,]$, we have:
\begin{equation}
T_1(t):=
\left\{
    \begin{tabular}{ll}
    $\displaystyle -\frac{T_\mathrm{surf}-T_\mathrm{bot.}}{H_\mathrm{bot}}\left(1-\frac{t}{\widehat{t}}\right)y
    -(T_\mathrm{surf}-T_\mathrm{fr})\frac{t}{\widehat{t}}
    +T_\mathrm{surf}$, & for $y\in[H_\mathrm{bot},H_\mathrm{fr}]$,   \\  [0.2cm]
    $\displaystyle -\frac{T_\mathrm{surf}-T_\mathrm{bot.}}{H_\mathrm{bot}}\left(1-\frac{t}{\widehat{t}}\right)y
    -\frac{T_\mathrm{surf}-T_\mathrm{fr}}{H_\mathrm{fr}}\frac{t}{\widehat{t}}\,y
    +T_\mathrm{surf}$, & for $y\in[H_\mathrm{fr},0]$,
\end{tabular}
\right.
\label{T1onGammaD}
\end{equation}
where also $T_\mathrm{fr}:=-15$ $^\circ$C. For $t>\widehat{t}$, we assume that $T_1$ remains fixed and is given by $T_1(\widehat{t})$ in (\ref{T1onGammaD}). The heat conduction in the modelled case is, hence, triggered by a significant difference of temperatures on $\Gamma_D$ and in $S$.

\begin{figure}%[h]
\begin{center}
\includegraphics[width=1.0\textwidth]{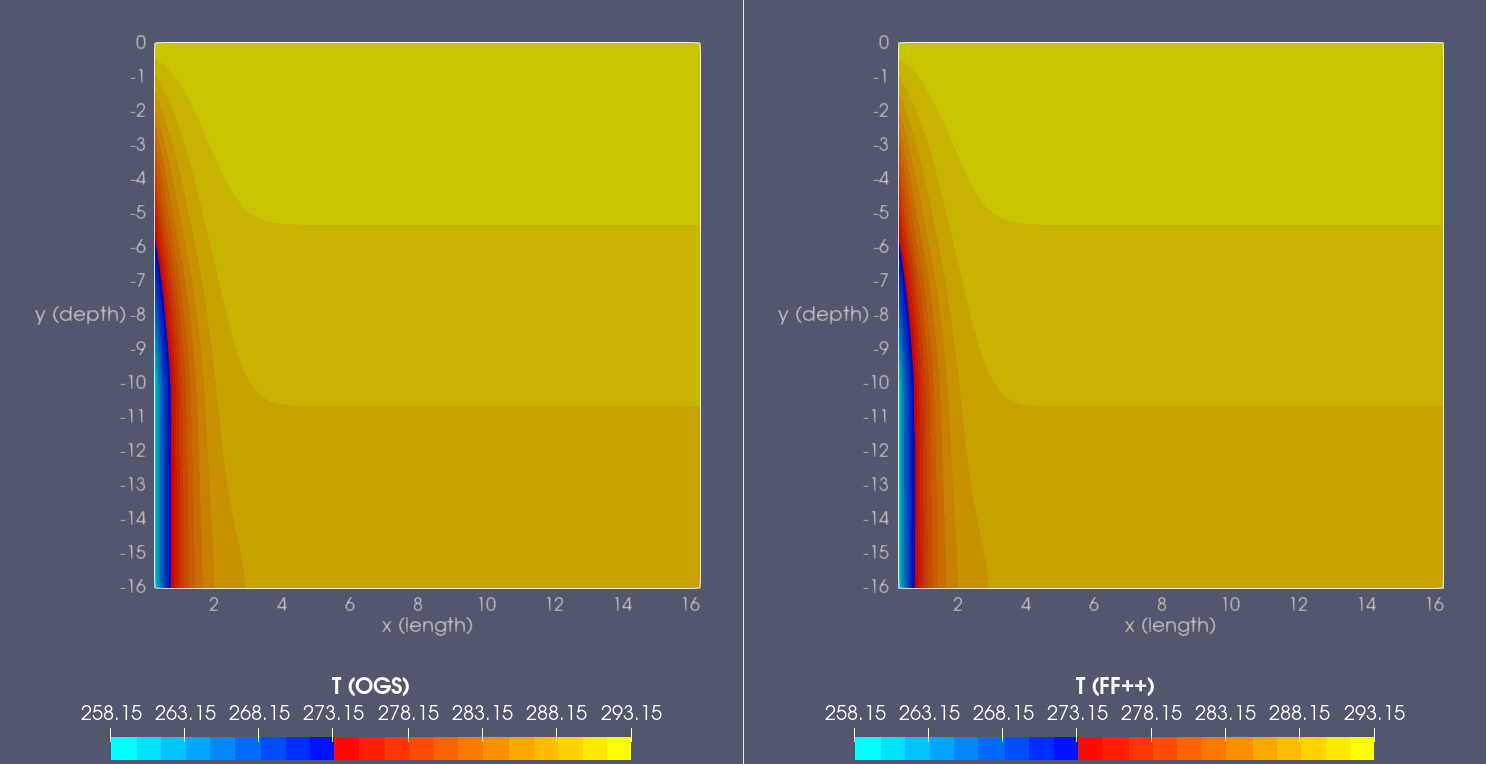}\\
\includegraphics[width=1.0\textwidth]{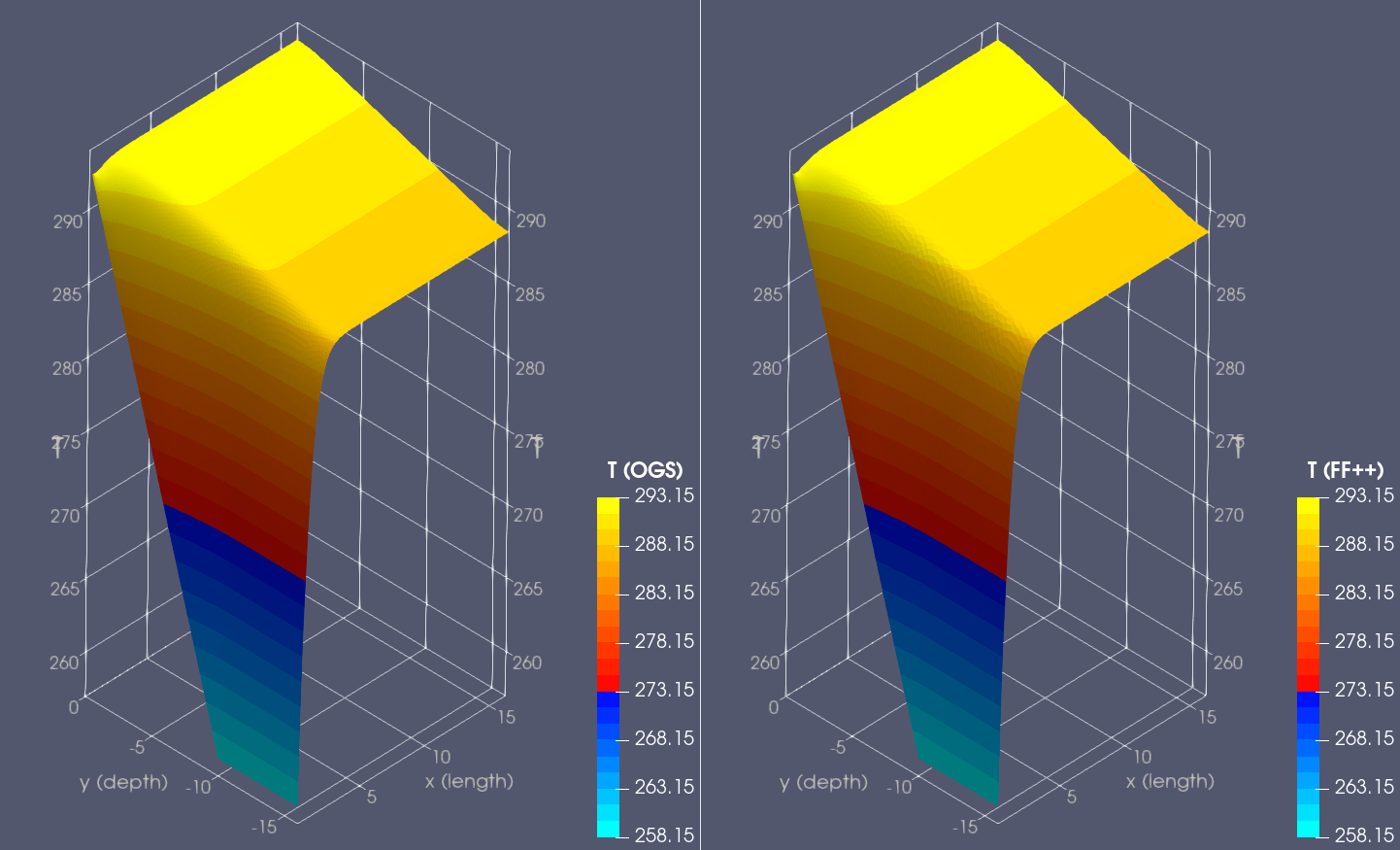}
\end{center}
\caption{Comparison of the simulation results obtained by OpenGeoSys-6 and FreeFem++ codes (in plots, termed OGS and FF++) for the setup from \autoref{fig_8}: temperature $T$ distribution within the block $S$ after 720 hours (30 days) of cooling, 2D and 3D views; temperature is given in the unit Kelvin.}
\label{fig_6}
\end{figure}

The results of modeling are depicted in \autoref{fig_6}, where we plot the temperature distribution in the soil block after $720$ hours (30 days) of cooling, and also compare the outcomes of the two simulation codes used: OGS and FF++. The color legend of $T$ in the corresponding ParaView plots is tuned such that the amount of ice formed around BHEs can be identified. As expected, ice formation occurs in the vicinity of $\Gamma_D$, more specifically, near the segment of $\Gamma_D$, where the negative temperature has been prescribed. In the rest of the domain, temperature distribution remains identical to the initial condition, as can also be expected.

In either case, the $P_1$-based approximations of $T$ were computed. The underlying meshes are generated with different software but in a way that the number of nodes (and elements) is comparable (almost identical). Also, each mesh is refined in a finite strip in the vicinity of $\Gamma_D$, where ice formation is expected. This is done to be able to resolve localization due to $S_\mathrm{I}$ and $\frac{\mathrm{d}S_\mathrm{I}}{\mathrm{d}T}$. The total simulation time interval is $t\in[0,2592000\,\mathrm{s}]=[0\,\mathrm{h},720\,\mathrm{h}]$, and the time-step increment $\Delta t:=900\,\mathrm{s}=15\,\mathrm{min}$.

\begin{figure}%[h]
\begin{center}
\includegraphics[width=0.85\textwidth]{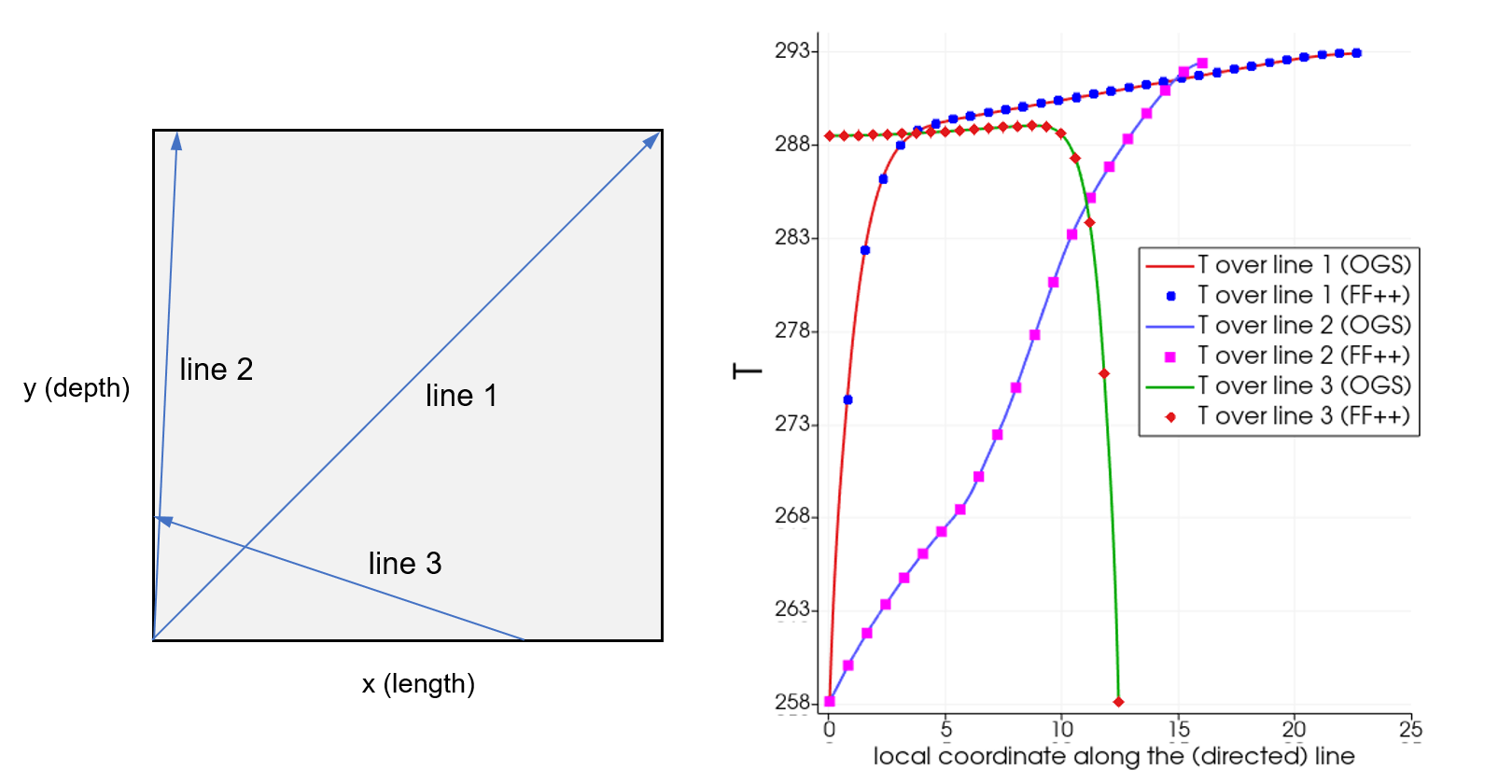}
\end{center}
\caption{Comparison of the results for temperature distribution from \autoref{fig_6} over the directed lines within $S$; origin of the horizontal axis on the right plot corresponds to line's origin.}
\label{fig_7}
\end{figure}

In \autoref{fig_6}, already the qualitative similarity of the OGS and FF++ results can be observed. \autoref{fig_7} presents the corresponding results from \autoref{fig_6} plotted over the three different (directed) lines within the domain $S$, thus also making the quantitative comparison feasible. For the selected lines, the compared data seems identical point-wise.

%%%%%%%%%%%%%%%%%%%%%%%%%%%%%
\subsection{Benchmark tests for the H+freezing problem: ice barrier for the liquid flow}
\label{subsec:BenchmarkH+freezing}

Herein, we examine the hydraulic component of our model given by the H+freezing equation (\ref{BalMass}). The goal is to simulate how the prescribed temperature variation (especially when it leads to ice formation within the domain of a specific intuitive shape) affects the initially steady laminar flow. 

{\bf Note on temperature-dependent hydraulic flow:} In the THM modeling of saturated porous media with phase change (herein, it is a liquid-to-ice phase transition), it proves necessary to make the hydraulic parameters present in the Darcy law (\ref{Darcy}) temperature-dependent. More precisely, the liquid-phase relative permeability $\kappa_\mathrm{rel}$ and the dynamic viscosity $\mu_\mathrm{LR}$ should depend on the ice-fraction and (thus) on the temperature in order to provide the decrease of liquid water flow in the region where ice forms.
A typical representation for $\mu_\mathrm{LR}(T)$ -- first introduced in \cite{Grant2000} and employed, e.\,g., in \cite{Coussy2005,Zhou2014,NaSun2017}, as well as in our case here -- reads 
\begin{equation}
    \mu_\mathrm{LR}=\mu_\mathrm{LR}(T):=\mu_\mathrm{LR}^0\,\delta(T) 
    \quad\text{with}\quad \delta(T):=1.5963\cdot 10^{-2}\exp\left(\frac{509.53}{T-150\,\mathrm{K}}\right)
\end{equation}
and $\mu_\mathrm{LR}^0$ being the reference value.
%$\mu_\mathrm{LR}=\mu_\mathrm{LR}(T):=\mu_\mathrm{LR}^0\delta(T)$  with $\mu_\mathrm{LR}^0$ being the reference value and $\delta(T):=1.5963\cdot 10^{-2}\exp\left(\frac{509.53}{T-150\,\mathrm{K}}\right)$.
For the permeability a drop by $b$ orders of magnitude is assumed if the matrix contains pore ice, thus enabling the required impact on the liquid flow.
This can be achieved by assuming $\kappa_\mathrm{rel}^\mathrm{L}=1$ when $T>T_\mathrm{fr}$ (liquid phase) and $\kappa_\mathrm{rel}^\mathrm{L}=10^{-b}$, $b>0$ when $T<T_\mathrm{fr}$ (ice phase). Such $b$ can be viewed as a user-defined model parameters to be calibrated. 
In our simulations, the ansatz 
\begin{equation}
    \kappa_\mathrm{rel}^\mathrm{L}=1-(1-10^{-b})S_\mathrm{I}(T)=
    \left\{
    \begin{tabular}{ll}
    $1$, & liquid phase,   \\  [0.2cm]
    $10^{-b}$, & ice phase,
    \end{tabular}
    \right.
\label{ourKappaRel}
\end{equation}
with $S_\mathrm{I}$ given by (\ref{Sigmoid}) and $b>0$ has been employed. 
The benchmark illustrating the use of the above $\kappa_\mathrm{rel}^\mathrm{L}$ from (\ref{ourKappaRel}) with $b=4$ is presented now. 

\begin{figure}[h!]
\begin{center}
\includegraphics[width=1.0\textwidth]{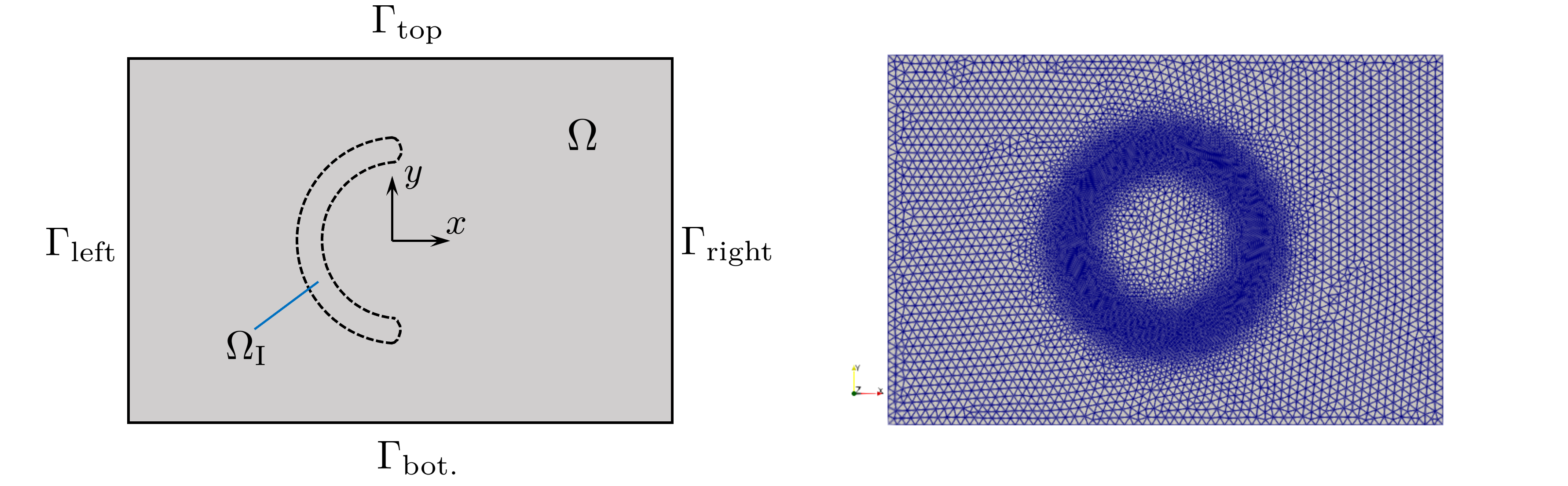}
\end{center}
\caption{A computational domain $\Omega$ initially occupied by pore liquid and a subdomain $\Omega_\mathrm{I}$ where ice formation due to the prescribed temperature evolution is expected (on the left); the finite element mesh (on the right).}
\label{fig_Hpr1}
\end{figure}

\begin{figure}[h!]
\begin{center}
\includegraphics[width=1.0\textwidth]{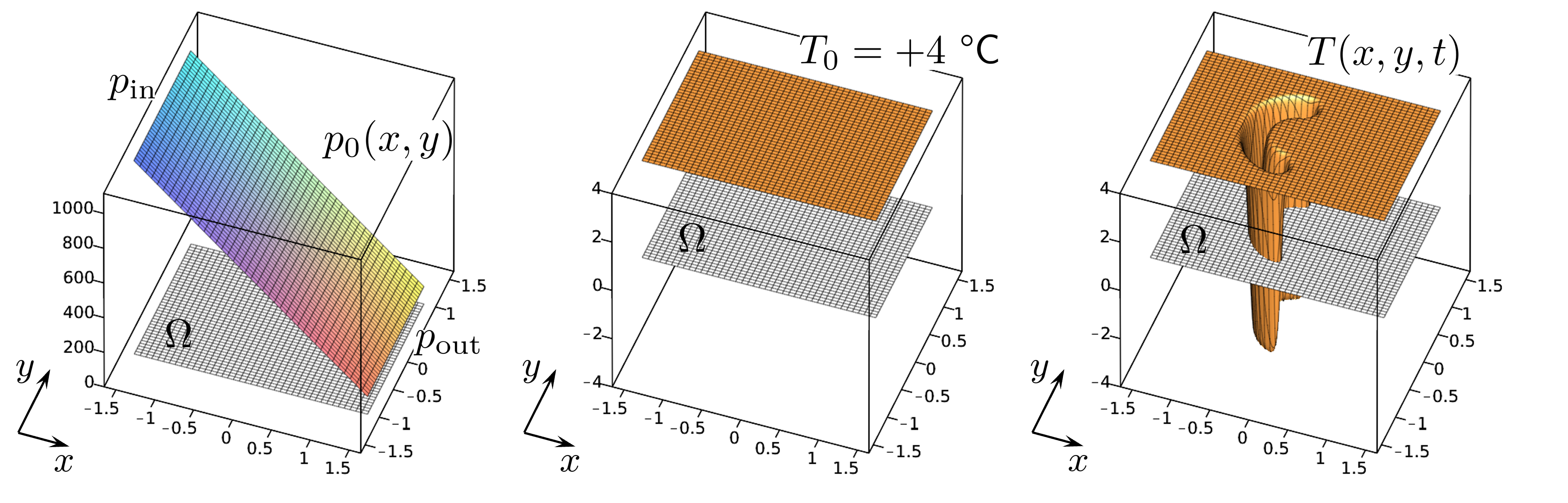}
\end{center}
\caption{Initial conditions for liquid pressure $p_\mathrm{LR}$ and temperature $T$ in $\Omega$ (left and middle, resp.); the profile of the applied temperature at a final time-step $t_\mathrm{fin}$ (on the right) which triggers freezing of liquid in $\Omega$ in such a way that the grown ice-body occupies a horse-shoe shaped subdomain $\Omega_\mathrm{I}$ in Figure \ref{fig_Hpr1}.}
\label{fig_Hpr2}
\end{figure}

Our computational domain is a rectangle $\Omega:=(-L_x,L_x)\times(-L_y,L_y)$ with $L_x=1.5$~m and $L_y=1$~m, see \autoref{fig_Hpr1}. As an initial TH state, we assume in $\Omega$, a spatially constant temperature $T_0(x,y):=+4$ °C and a linearly varying pressure
\begin{equation*}
    p_0(x,y):=-\frac{1}{2L_x}(p_\mathrm{in}-p_\mathrm{out})x
    +\frac{1}{2}(p_\mathrm{in}+p_\mathrm{out}),
\end{equation*}
where $p_\mathrm{in}=11 p_\mathrm{atm}$ and $p_\mathrm{out}=p_\mathrm{atm}=101325$ Pa are prescribed on $\Gamma_\mathrm{left}$ and $\Gamma_\mathrm{right}$, respectively. We depict $p_0$ and $T_0$ in Figure \autoref{fig_Hpr2}. These conditions provide an initially steady laminar liquid flow within the domain.

In our H+freezing formulation to be solved for the unknown pressure $p$, we prescribe temperature evolution $T(x,y,t)$ in such a way that it triggers a growth of an ice-body in $\Omega_\mathrm{I}\subset\Omega$, with $\Omega_\mathrm{I}$ being of some interesting shape. In our case, we opted for $\Omega_\mathrm{I}$ to be horse-shoe alike, see Figure \ref{fig_Hpr1} (left) for the plot of $T$ and Figure \ref{fig_Hpr2} (right) for the plot of $\Omega_\mathrm{I}$. The idea and our expectation is that the created ice-barrier will divert the flow such that in a shielded area behind the ice-barrier the flow will cease (almost) completely. We should be able to see this effect once we recover the corresponding filter velocity $\widetilde{\bm w}_\mathrm{LS}$ from the computed pressure variable $p$. 
The impact is assessed by calculating the Darcy filter velocity $\widetilde{\bm w}_\mathrm{LS}$: one observes that a liquid flow in the horse-shoe region $\Omega_\mathrm{I}\subset\Omega$ occupied by the grown ice-body ceases. Note that, in this case, $\widetilde{\bm w}_\mathrm{LS}=0$ in $\Omega_\mathrm{I}$ due $\kappa_\mathrm{rel}^\mathrm{L}\ll 1$ is, clearly, in contrast to $\widetilde{\bm w}_\mathrm{LS}=0$ observed behind the domain $\Omega_\mathrm{I}$, since the latter was caused by the condition $\nabla p=0$.

Notice that the equation of $T$ is lengthy and cumbersome. However, for the sake of completeness, it is  presented herein. Due to the envisioned shape of $\Omega_\mathrm{I}$ and location, it proves convenient to use polar coordinates $r=(x^2+y^2)^\frac{1}{2}$ and $\theta$. The desired $T$ reads: 

\begin{equation}
T(x,y,t):=
\left\{
    \begin{tabular}{ll}
    $T_0$, & for $t\in(0,t_1]$,   \\  [0.2cm]
    $\displaystyle T_0-(T_0-T_\mathrm{fin})\frac{t-t_1}{t_2-t_1}\exp{\left(-\frac{(r-R)^2}{a^2}\right)}q(\theta)$, & for $t\in(t_1,t_2]$, \\  [0.2cm]
    $\displaystyle T_0-(T_0-T_\mathrm{fin})\exp{\left(-\frac{(r-R)^2}{a^2}\right)}q(\theta)$, & for $t\in(t_2,t_\mathrm{fin}]$,
\end{tabular}
\right.
\label{TforHorseShoe}
\end{equation}
where $T_\mathrm{fin}=-4$ °C, $\{t_1,t_2,t_\mathrm{fin}\}=\{8,12,20\}$ in hours, $\{R,a\}=\{0.5,0.08\}$ and
\begin{equation*}
q(\theta):=\frac{1+e^{h(\theta-\pi/2)}+e^{2h\theta}}{1+e^{h(\theta-\pi/2)}+e^{h(\theta+\pi/2)}+e^{2h\theta}},
\end{equation*}
with $h=20$. Note that $q(\theta)$ provides that the horse-shoe domain triggered by $T$ is 'half-circular' in the $\theta$-direction, whereas the $\exp{\left(-(r-R)^2/a^2\right)}$ component of the above $T$ models spatial 'localization' of the domain in the $r$-direction.

As can be guessed from (\ref{TforHorseShoe}), our anticipated total process time is 12 hours, while the 'active freezing' phase (the decrease in temperature from $+4$ °C to $-4$ °C inside of $\Omega_\mathrm{I}$) occurs within the time interval between 4 and 8 hours from the start. In calculations, we chose 3 minutes to be our time-step increment. The finite element mesh is depicted in Figure \ref{fig_Hpr1}, right. It has been refined -- again, for the sake of accurate resolution of $T$ and related $S_\mathrm{I}$ -- in the annulus containing the prospective ice-body domain $\Omega_\mathrm{I}$. The calculated pressure $p$ is approximated by using the $P_1$-triangles. Finally, the material properties data used in this benchmark are presented in the following table.

\begin{table}[h]
\small
\centering
\begin{tabular}{ l | l | l  }
	solid phase & liquid phase & ice phase \\
	\hline
	$\varrho_\mathrm{SR}=2000$ kg/m$^3$ &  $\varrho_\mathrm{LR}=1000$ kg/m$^3$  &  $\varrho_\mathrm{IR}=920$ kg/m$^3$ \\
	$K_\mathrm{SR}=1.67\cdot 10^{10}$ Pa &  $K_\mathrm{LR}=2.2\cdot 10^{9}$ Pa  &  $K_\mathrm{IR}=7.81\cdot 10^{9}$ Pa \\
	$\alpha_T^\mathrm{S}=1.2\cdot 10^{-5}$ 1/K &  $\alpha_T^\mathrm{L}=0.7\cdot 10^{-4}$ 1/K  &  $\alpha_T^\mathrm{I}=0.55\cdot 10^{-4}$ 1/K \\
	 & $\mu_\mathrm{LR}^0=1.8\cdot 10^{-3}$ Pa s & $\ell=3.34\cdot 10^5$ J/kg \\
	\hline
	\multicolumn{3}{l}{Intrinsic permeability, $\kappa_\mathrm{i}=0.8\cdot 10^{-15}$ m$^2$}\\
    \multicolumn{3}{l}{Coefficient $b$ in (\ref{ourKappaRel}), $b=4$} \\
    \multicolumn{3}{l}{Porosity, $\phi=0.35$}\\
	\multicolumn{3}{l}{Sigmoid function $S_\mathrm{I}$ coefficient, $k=2$}\\
	\multicolumn{3}{l}{Freezing temperature, $T_\mathrm{fr}=0$ $^\circ$C (273.15 K)}\\
	\hline
\end{tabular}
\caption{Material properties and parameters for the H+freezing benchmark.}
\label{table_MatData_Hfreezing}
\end{table}

\begin{figure}[h!]
\begin{center}
\includegraphics[width=1.0\textwidth]{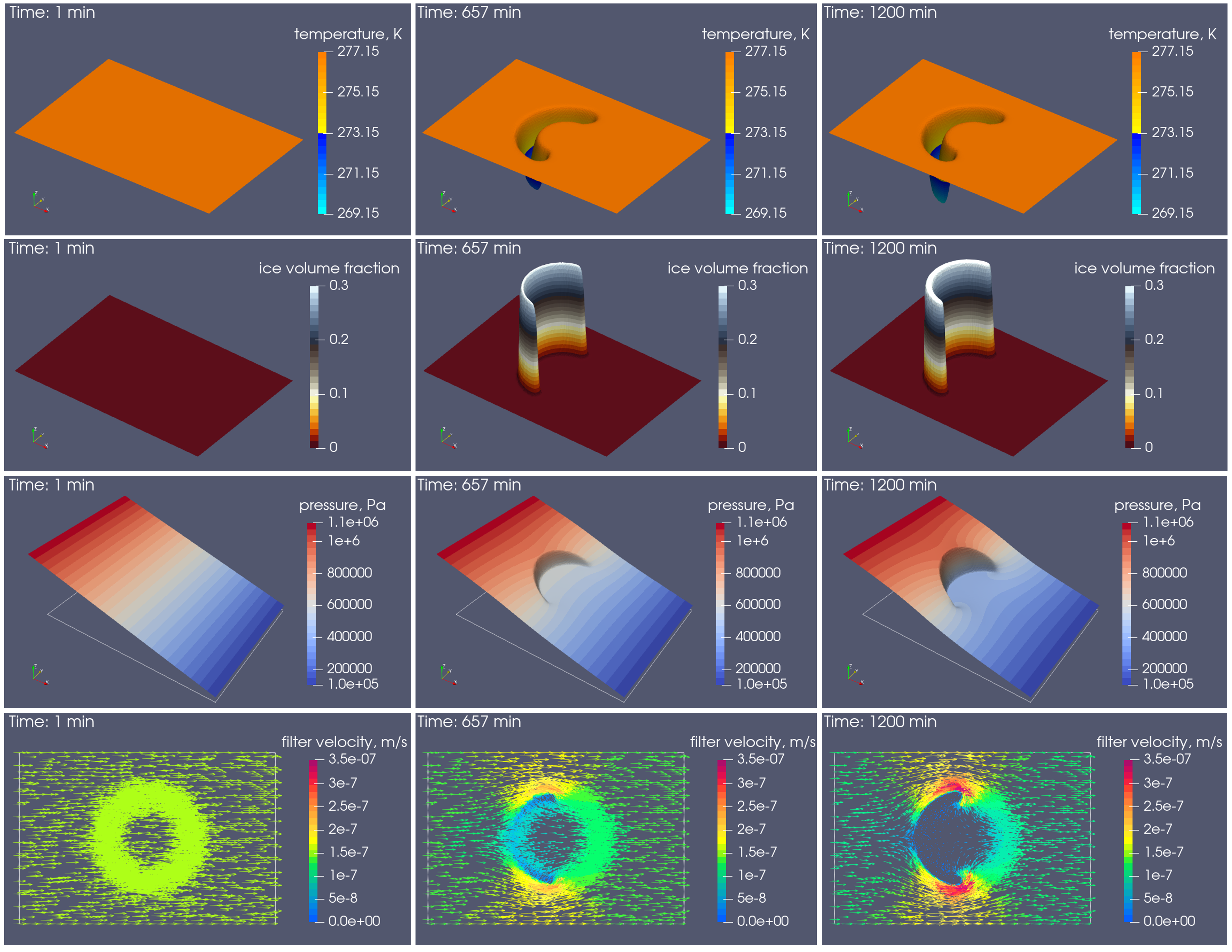}
\end{center}
\caption{Applied temperature $T$ (the prescribed quantity), the induced ice volume fraction $\phi_\mathrm{I}=\phi S_\mathrm{I}(T)$, calculated pressure $p$ and related filter velocity $\widetilde{\bm w}_\mathrm{LS}$ in $\Omega$ prior to freezing (1 min.), at the onset of freezing (657 min.) and at a final time step (1200 min.) when freezing has been finalized.}
\label{fig_Hpr3}
\end{figure}

The results of our simulations are shown in Figure \ref{fig_Hpr3}. We present snapshots of the prescribed $T$, the induced ice-volume fraction $\phi_\mathrm{I}=\phi S_\mathrm{I}(T)$, the calculated $p$ and related filter velocity (in this case, liquid) at the initial unfrozen state, during freezing, and after freezing has been finalized and a new stead-state has occurred.

It can be seen that the horse-shoe ice-body shields the corresponding area downstream, where the flow ceases completely due to the zero pressure gradient (notice that $p$ is indeed flat there). This was expected because of the experiment design. It can also be noticed that the flow velocity between the ice-body and the horizontal boundaries ($\Gamma_\mathrm{top}$ and $\Gamma_\mathrm{bot.}$ in Figure \ref{fig_Hpr1}) increased relative to the initial velocity and the velocity observed in the wider "channels" of $\Omega$ upstream and downstream the barrier. This agrees well with the theory.

%%%%%%%%%%%%%%%%%%%%%%%%%%%%%
\subsection{Benchmark tests for the M+freezing problem}
\label{subsec:BenchmarkM+freezing}
The following section finalizes the verification of our OGS implementation for the THM+freezing model. It regards the mechanical (M+freezing) part of the formulation which is thoroughly tested.  More specifically, we design and compute the two benchmarks, which enable us to evaluate the 'performance' of both the ice-related ingredient of the stress-strain relation, as well as the stress-stress relation overall.

%%%%%%%%%%%
\subsubsection{Verifying the 9\% expansion due to water-to-ice phase change}

The aim of this benchmark is to verify that our model is capable of simulating the 9$\%$ volumetric expansion during the liquid-to-ice phase transition. As this must be captured by the $\bm\sigma_\mathrm{I}$-stress component of formulation (\ref{sigESI_3}), the material properties in the benchmark are chosen such that $E_\mathrm{S}\ll E_\mathrm{IR}$. In this way, and also with the absence of mechanical loading, the deformation of a specimen -- with $\alpha_{\phi_\mathrm{I}}=0.03$ being the strain increase in each space direction -- will occur solely due to liquid freezing. 

\begin{figure}[h!]
\begin{center}
\includegraphics[width=1.0\textwidth]{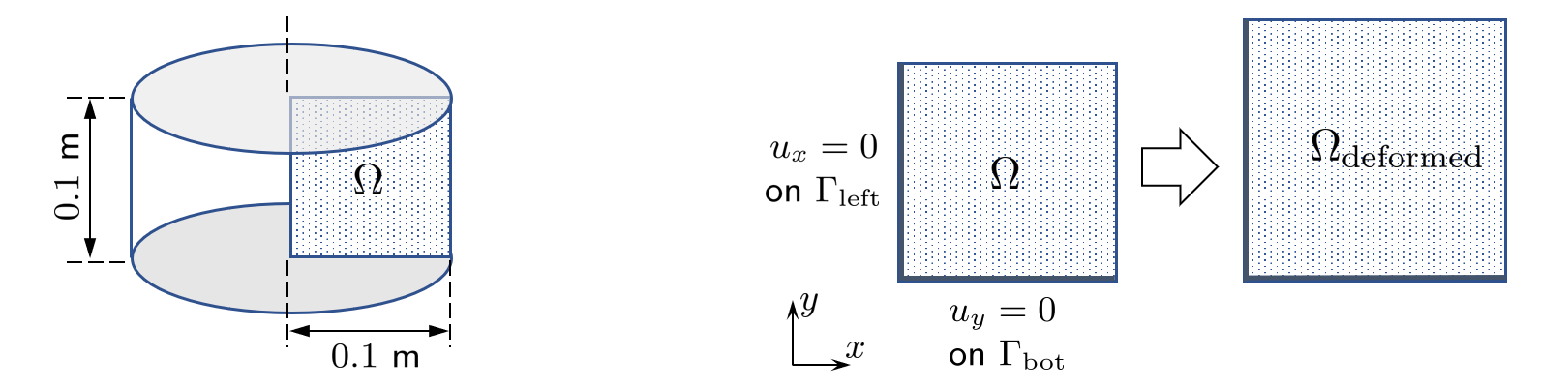}
\end{center}
\caption{Fully saturated column expansion due to water-to-ice phase transition: geometry (on the left) and the 2D computational setup along with the expected deformed configuration (on the right).}
\label{fig_Ice_strain_1}
\end{figure}

\begin{figure}[h!]
\begin{center}
\includegraphics[width=1.0\textwidth]{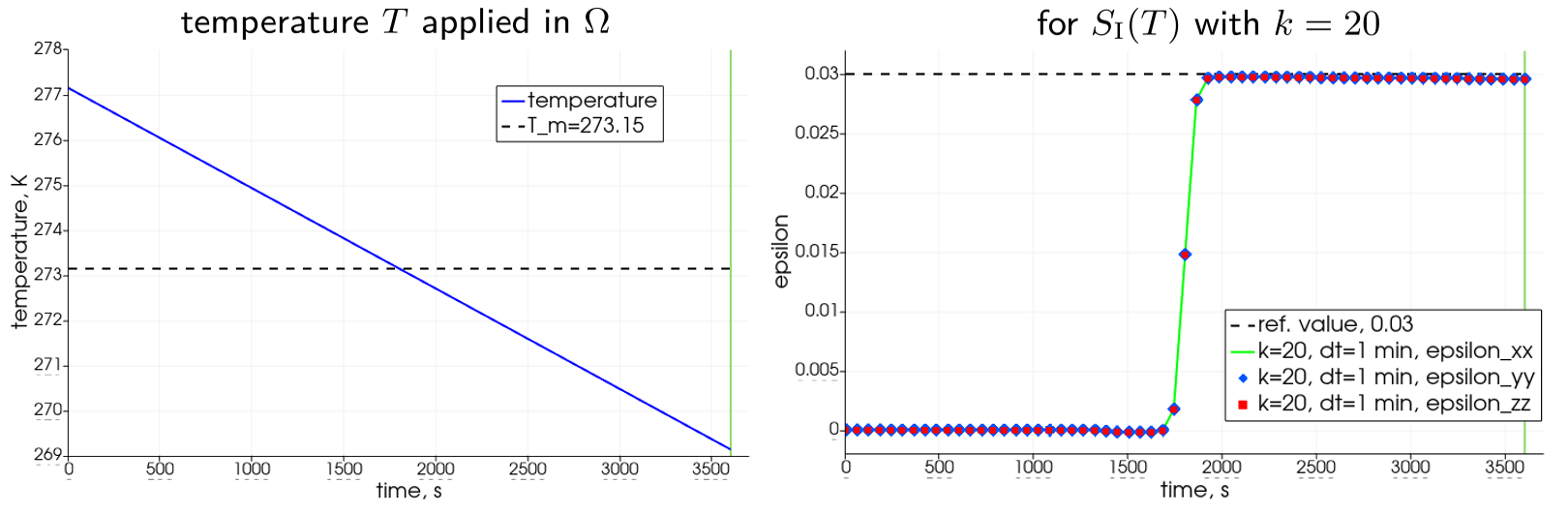}
\end{center}
\caption{The prescribed temperature loading applied to the specimen (on the left) and the induced normal strain evolution due to phase change (on the right).}
\label{fig_Ice_strain_2}
\end{figure}

For the geometric setup we consider a fully saturated cylindrical column whose bottom edge is supported by a rigid foundation. Using the axial symmetry, the problem is reduced from 3 to 2 dimensions with a simple (square) computational domain and related (Dirichlet) boundary conditions, see \autoref{fig_Ice_strain_1}, where an expected deformed configuration is also sketched. Again, we apply no mechanical loading to the specimen, whereas the thermal loading is presented by the temperature evolution over time. The temperature $T$ is prescribed as a constant in $\Omega$ at each time step and decays linearly from $+4$ $^\circ$C to $-4$ $^\circ$C during one hour, as depicted in \autoref{fig_Ice_strain_2}, left.

Material data used in the computations are presented in Table \ref{table_MatData_Mfreezing}. Note that no liquid phase is present since we do not solve the hydraulic equation and the temperature field is prescribed. The two parameters that are varied are the time step increment $\Delta t$ (also denoted as $\mathrm{dt}$ in the corresponding captions) and the parameter $k>0$ in the Sigmoid function $S_\mathrm{I}$ which governs the width of a temperature-related phase transition zone. More specifically, we take $\Delta t\in\{10\,\mathrm{s}, 30\,\mathrm{s}, 1\,\mathrm{min}\}$ and $k\in\{2,5,20,50\}$. The domain $\Omega$ is descretized with only one element (which is possible here since the homogeneous setup implies no freezing front propagation within the domain). %Finally, the FE approximation of the components of strain tensor $\bm\varepsilon$ uses the $Q_1$-quadrilaterals.

\begin{table}[h]
\small
\centering
\begin{tabular}{ l | l  }
	solid matrix/skeleton & ice phase \\
	\hline
%	$\varrho_\mathrm{SR}=2000$ kg/m$^3$ &  $\varrho_\mathrm{IR}=920$ kg/m$^3$ \\
%	$c_{p\mathrm{S}}=900$ J/(kg K) &  $c_{p\mathrm{I}}=2090$ J/(kg K) \\
%	$\lambda_\mathrm{SR}=1.1$ W/(m K) &  $\lambda_\mathrm{IR}=2.2$ W/(m K) \\
%	   & $\ell=3.34\cdot 10^5$ J/kg \\
%    \hline
    $\alpha_T^\mathrm{S}=1.2\cdot 10^{-5}$ 1/K &  $\alpha_T^\mathrm{I}=5.5\cdot 10^{-5}$ 1/K \\
    & $\alpha_{\phi_\mathrm{I}}=0.03$ \\
    \hline
%    solid matrix/skeleton parameters: & \\
    $E_\mathrm{S}=30$ Pa & $E_\mathrm{IR}=10$ GPa \\
    $\nu_\mathrm{S}=0.2$ Pa & $\nu_\mathrm{IR}=0.2$ \\
	\hline
	\multicolumn{2}{l}{Porosity, $\phi=0.35$}\\
	\multicolumn{2}{l}{Melting temperature, $T_\mathrm{fr}=0$ $^\circ$C (273.15 K)}\\
	\hline
\end{tabular}
\caption{Material properties and parameters.}
\label{table_MatData_Mfreezing}
\end{table}

\autoref{fig_Ice_strain_2}, right, depicts the evolution of strain tensor components $\{\varepsilon_{xx},\varepsilon_{yy},\varepsilon_{zz}\}$ computed for the fixed parametric pair $(\Delta t,k)=(1\,\mathrm{min},20)$. All three strains behave identically. More importantly, as expected, they transit from 0 to the reference magnitude of 0.03 during freezing, in accordance to the term $\alpha_{\phi_\mathrm{I}}S_\mathrm{I}(T)$. 

\begin{figure}[h!]
\begin{center}
\includegraphics[width=1.0\textwidth]{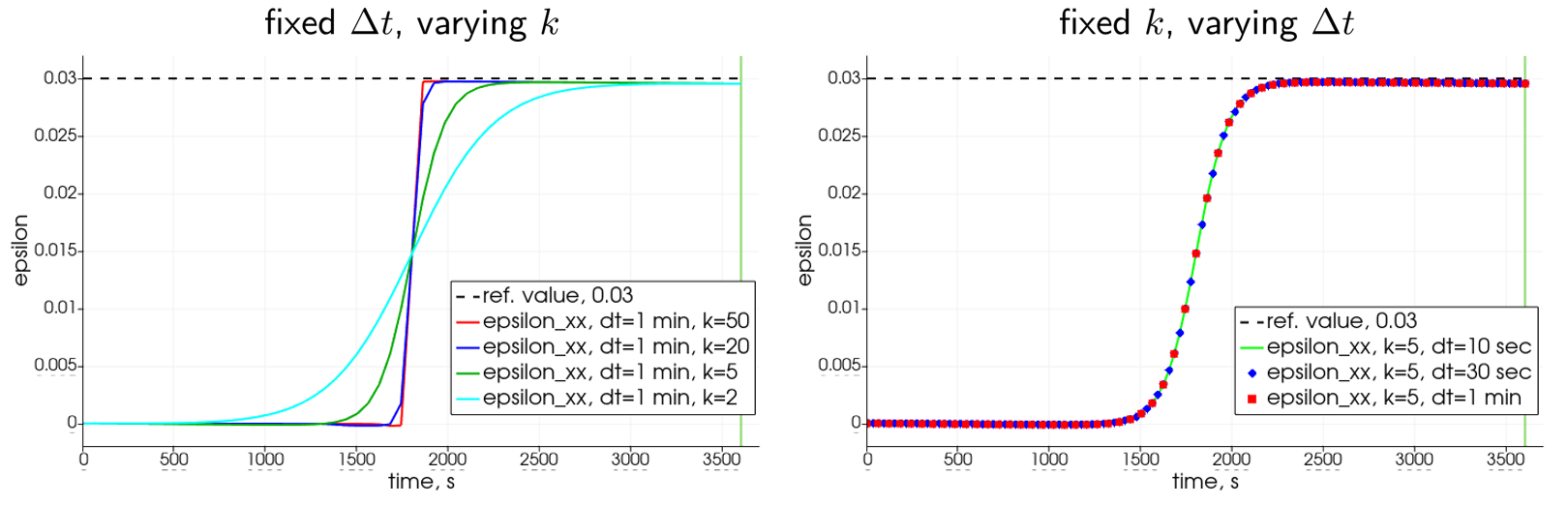}
\end{center}
\caption{Parametric studies for the computed volumetric expansion in dependence on the time step and parameter $k$ in $S_\mathrm{I}(T)$ governing the thickness of a phase transition zone.}
\label{fig_Ice_strain_3}
\end{figure}

\autoref{fig_Ice_strain_3} details the parametric studies for the computed $\varepsilon_{xx}$: on the left plot, for the fixed time increment, one observes that for any considered $k$ the corresponding strains transit up to the required value 0.03 and, as expected, the increase of the parameter yields a steeper and more localized transition zone, almost mimicking the Heaviside-like behavior at $k=50$. It is interesting to observe a slight downward deviation of $\varepsilon_{xx}$ from the horizontal reference line in the post-freezing time range (that is, when the prescribed temperature keeps on decreasing from $T_\mathrm{fr}$ to $-4$ $^\circ$C). This behavior is physical and induced by the ice contraction in this temperature range. The right plot of \autoref{fig_Ice_strain_3} presents the evolution of $\varepsilon_{xx}$ -- specifically, the required transit during the phase change -- for a fixed steepness-related parameter $k$ and varying time step size. All three computational results seem identical. 

%%%%%%%%%%%
\subsubsection{Deformation and freezing of a poro-elastic column}

In this example, we consider a fully-saturated poro-elastic column which is subject to a combination of thermal and mechanical loading. This loading is prescribed in a way that the specimen passes various stages of deformation: purely mechanical deformation of the solid matrix, deformation of the solid-ice mixture, as well as deformation induced by the liquid-to-ice phase transition. We thus check the plausibility of the M+freezing model given by the IBVP for equation (\ref{BalMom}).

\begin{figure}[h!]
\begin{center}
\includegraphics[width=1.0\textwidth]{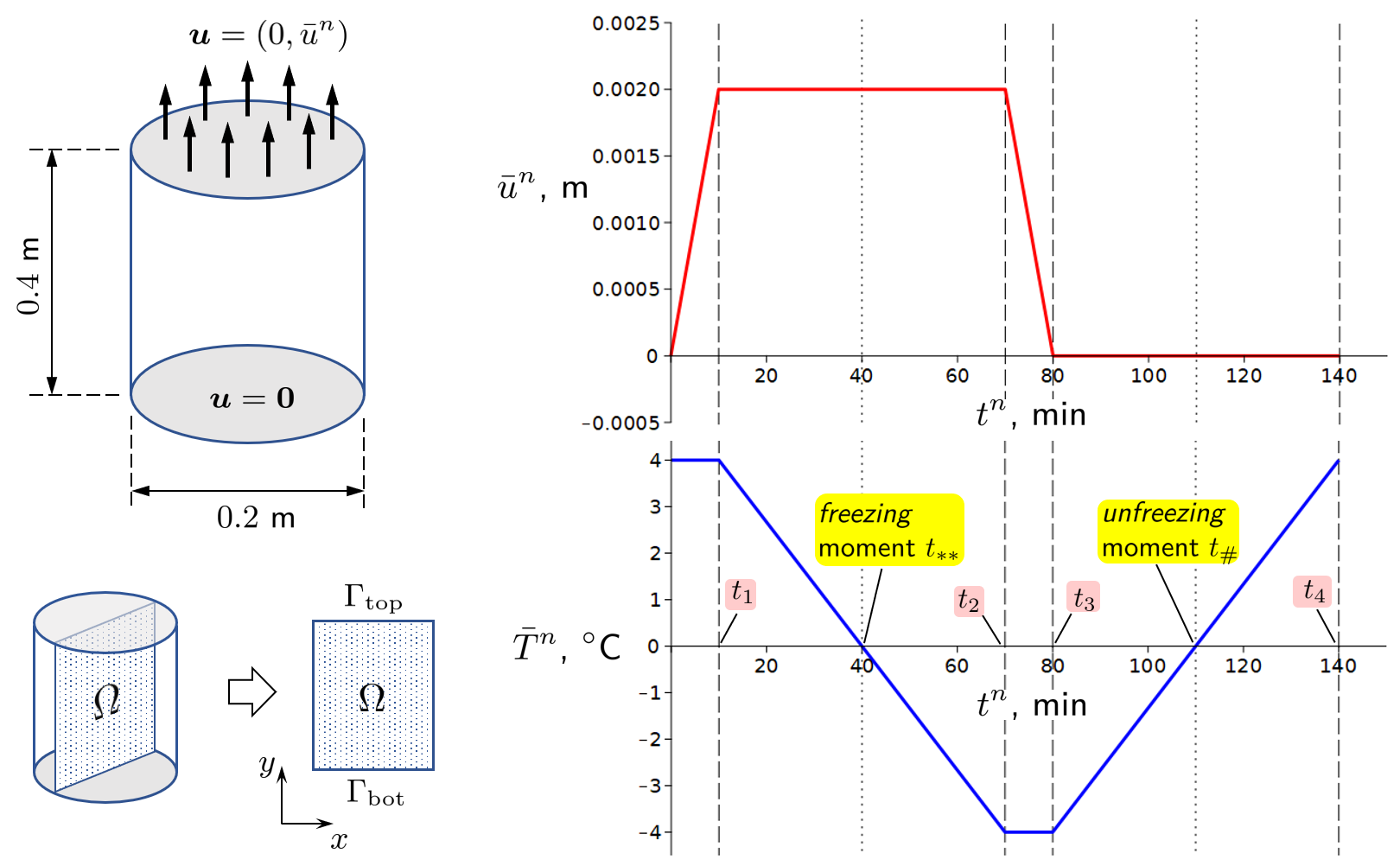}
\end{center}
\caption{Fully saturated poro-elastic column geometry and mechanical loading (on the left); plots of the prescribed displacement and thermal loading applied to the specimen (on the right).}
\label{fig_Column_1}
\end{figure}

\autoref{fig_Column_1}, left, depicts the geometric and mechanical loading setup: the cylindrical column is fixed at the bottom edge and the incremental vertical displacement loading $\bar{u}^n$, with $n\geq1$ being a time step, is applied at the column's top boundary. Simultaneously, the incremental thermal loading $\bar{T}^n$ is applied within the column. This implies, that all temperature dependent coefficients in the momentum balance equation are varied by setting $T=\bar{T}^n$ therein. \autoref{fig_Column_1}, right, details both $\bar{u}^n$ and $\bar{T}^n$ including the time intervals of interest. The unknown we solve for is the displacement field $\bm u$. Notice that using the symmetry of the three-dimensional domain, we effectively consider and solve the two-dimensional problem in a diametrical cross-section, where $\Omega$ is our computational domain and $\Gamma_\mathrm{top}$ and $\Gamma_\mathrm{bot}$ are the corresponding edges. 

The material parameters used in the simulations are identical to those depicted in Table~\ref{table_MatData_Mfreezing} except for the solid matrix Young's modulus $E_\mathrm{S}$. In the present case, it is taken as $30$ GPa, replacing the original table value of $30$ Pa. Finite element discretization of $\Omega$ uses 10$\times$20 elements, and all components of $\bm\varepsilon$ are approximated by the $Q_1$-quadrilaterals.\footnote{In fact, similarly to the previous benchmark, the mesh here can be as coarse as possible too, since the setup implies no freezing front propagation within the domain.}

\begin{figure}[h!]
\begin{center}
\includegraphics[width=1.0\textwidth]{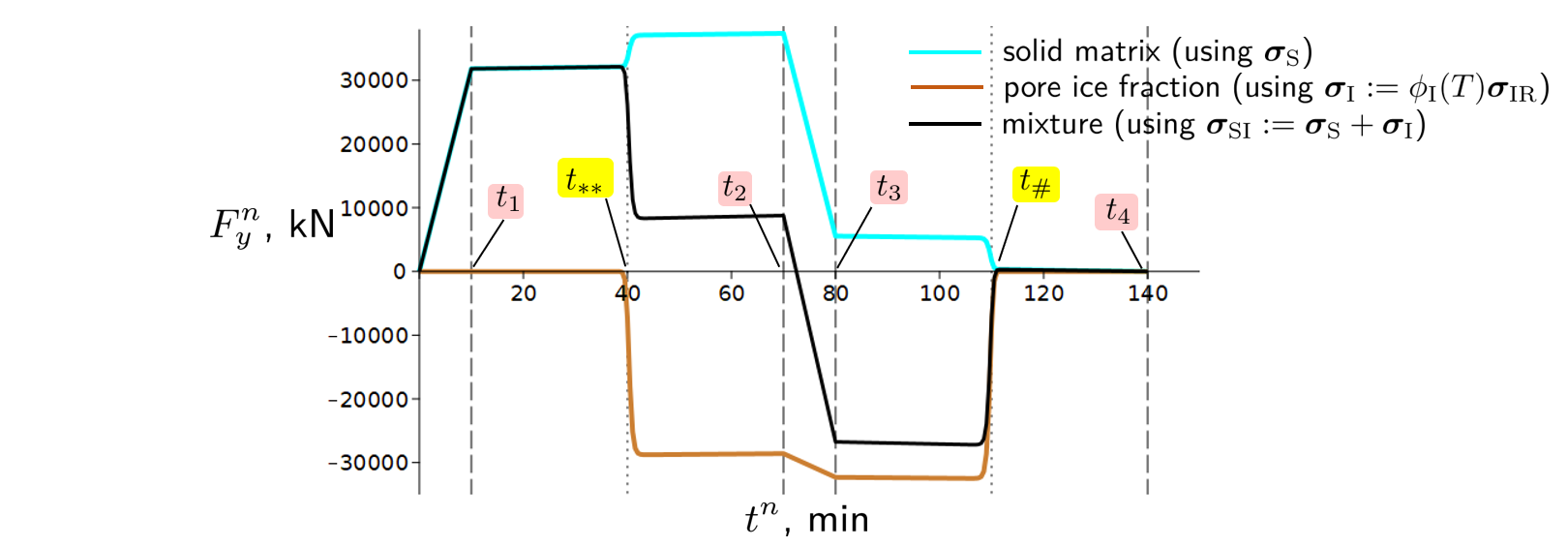}
\end{center}
\caption{Plots of the vertical reaction force $F_y^n$ for the solid matrix, the pore-ice and the solid-ice mixture.}
\label{fig_Column_2}
\end{figure}

To assess and analyze our simulation results, we calculate and record the vertical component $F_y$ of the reaction force on the top boundary $\Gamma_\mathrm{top}$ at each time step, i.\,e.
\begin{equation}
    \bm F^n=(F^n_x,F^n_y):=\int_{\Gamma_\mathrm{top}}
\bm{\sigma}(\bm u^n)\cdot\bm n\,
\mathrm{d}s, \quad \bm{\sigma}\in\{\bm{\sigma}_\mathrm{S},\bm{\sigma}_\mathrm{I},\bm{\sigma}_\mathrm{SI}\},
\label{reaction}
\end{equation}
where $\bm u^n:\Omega\rightarrow\mathbb{R}^2$ is the computed displacement solution vector, $\bm n$ is an outward normal on $\Gamma_\mathrm{top}$ and $\bm{\sigma}_\mathrm{S}$, $\bm{\sigma}_\mathrm{I}:=\phi_\mathrm{I}\bm\sigma_\mathrm{IR}$ and $\bm{\sigma}_\mathrm{SI}:=\bm{\sigma}_\mathrm{S}+\bm{\sigma}_\mathrm{I}$ are stresses related to the solid matrix, the pore ice fraction and the solid-ice mixture, respectively, see Eq. \eqref{sigESI_2}. The corresponding results are presented in \autoref{fig_Column_2}. 

Using both time-loading and time-reaction curves plotted in \autoref{fig_Column_1} and \autoref{fig_Column_2}, respectively, the following is observed:
\begin{enumerate}
    \item Time interval $(0,t_1)$: The solid matrix is under the  constant positive temperature and is experiencing vertical tension. The recovered reaction $F_y$ on $\Gamma_\mathrm{top}$ is linearly growing.
    \item Time interval $(t_1,t_2)$. The displacement loading on $\Gamma_\mathrm{top}$ is kept constant, whereas the specimen temperature is decreasing from positive to subzero with $t_{**}:=\frac{1}{2}(t_2-t_1)\in(t_1,t_2)$ being the freezing moment. 
    \begin{itemize}
        \item On sub-interval $(t_1,t_{**})$ the temperature is decreasing from the positive to $T_\mathrm{fr}$. The solid matrix contracts yielding the slight increase of the reaction $F_y$ on $\Gamma_\mathrm{top}$, as expected.
        \item At $t=t_{**}$, one has $T=T_\mathrm{fr}$ such that the Sigmoid ice-fraction indicator function takes the value $\frac{1}{2}$ (both liquid water and ice fractions are present equally). In other words, liquid-to-ice phase change happens in the "small vicinity" of $t_{**}$. Since during such phase transition the liquid expands by 9\% thus pushing up the fixed top edge, the reaction on $\Gamma_\mathrm{top}$ is expected to drop. It can be seen that the model fulfills such prediction. Also notice that after $t_{**}$, our specimen is already a mixture of a (deformed) solid matrix and an (undeformed) pore ice. 
        \item On sub-interval $(t_{**},t_2)$, the temperature keeps on decreasing thus causing further contraction of solid matrix and the already formed ice. This results in a slight linear increase of the reaction $F_y$ on $\Gamma_\mathrm{top}$.
    \end{itemize}
    \item Time interval $(t_2,t_3)$: The subzero temperature is kept constant, and we start unloading the completely frozen specimen by reducing the vertical displacement loading to its initial (zero) value. The actual unloading applies to the solid matrix, as it was previously deformed (stretched) vertically, whereas the pore ice is experiencing compression. This can be observed by assessing the recovered reaction curves of the corresponding constituents. The total reaction force on $\Gamma_\mathrm{top}$ also decays on this time interval and even enters the negative range. The latter is the indication that the solid-ice mixture remains under compression even when $\bar{u}^n$ on $\Gamma_\mathrm{top}$ reaches $0$. This is expected since for the pore ice this is  a deformed configuration.
    \item Time interval $(t_3,t_4)$: The vertical displacement on $\Gamma_\mathrm{top}$ is kept at zero level and we start warming the completely frozen specimen. The temperature is increased from negative to the initial positive value with $t_{\#}:=\frac{1}{2}(t_4-t_3)\in(t_4,t_3)$ being the melting moment.
    \begin{itemize}
        \item On sub-interval $(t_3,t_\#)$ the specimen expands due to the increase of thermal loading thus pushing upwards the fixed top boundary $\Gamma_\mathrm{top}$. The recovered reaction $F_y$ is hence slightly increasing in negative direction.
        \item In the "small vicinity" of melting moment $t_\#$ ice-to-liquid phase transition happens. This is accompanied by 9$\%$ of volume contraction of ice. Since the ice phase ceases to be present, the negative reaction force $F_y$ on $\Gamma_\mathrm{top}$ drops to a very small positive value. This value is not zero since a slight contraction of the solid matrix at temperature $T=T_\mathrm{fr}$ is still present.
        \item On sub-interval $(t_\#,t_4)$, further warming the solid matrix goes on till the temperature reaches the initial positive value. The specimen continues its thermal expansion and finally returns to its initial undeformed state. The reaction $F_y$ returns back to $0$ as well, as expected.
    \end{itemize}
\end{enumerate} 

\begin{figure}[h!]
\begin{center}
\includegraphics[width=1.0\textwidth]{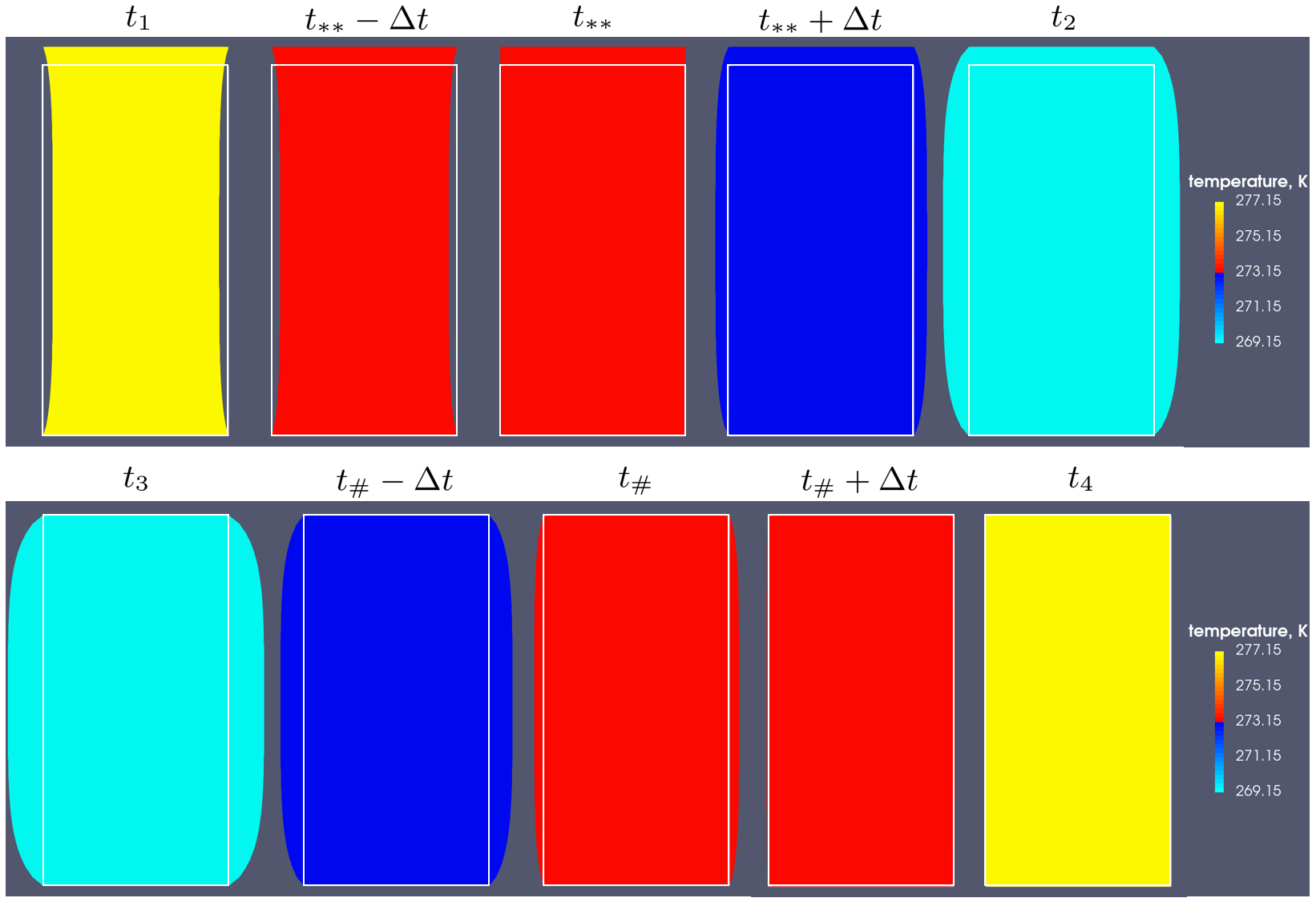}
\end{center}
\caption{Specimen deformation plots at the corresponding time moments; the computed displacement components $u_x$ and $u_y$ are exaggerated by factors 10 and 80, respectively.}
\label{fig_Column_3}
\end{figure}

Figure \ref{fig_Column_3} presents snapshots of the specimen in the deformed configuration at the time moments from Figures \ref{fig_Column_1} and \ref{fig_Column_2}. The fill corresponds to specimen's temperature, and a white frame stands for the undeformed configuration to highlight deformation shape and magnitude. (We notice that in each plot the computed displacement coefficients $u_x$ and $u_y$ have been exaggerated by factors 10 and 80, respectively). The deformation plots supplement and illustrate the above observations and conclusions drawn.

%%%%%%%%%%%%%%%%%%%%%%%%%%%%%%%%%%%%%%%%%%%%%%%%%%%%%%%%%%%%%%%%%%%%%%%%%%%%%%%%%%%%%%%%%%%%%%%%%%%%%%%%%%%%%%%%%
\section{Summary and Outlook}\label{sec:summary}

In this paper, the presented THM+F formulation was implemented in the FEM simulator OGS-6. In doing so, potential sources of unclarity have been addressed. Several numerical ingredients that have a qualitative and quantitative impact on the results, in particular the choice of sigmoid steepness parameter $k$ and the convexity of the underlying energy functional, were identified and analyzed as well.
The fully-coupled system of partial differential equations (THM+F) contains the T+freezing, H+freezing, and M+freezing sub-problems. A series of benchmark problems that examines each component of the formulation and the corresponding OGS-6 implementation was designed carefully.

Considering the T+freezing sub-problem, we showed that our implementation is capable of solving a one-dimensional Stefan melting problem resembling an analytical solution. A proper two-dimensional code verification was then performed using the method of manufactured solutions. Next, the simulation of soil freezing around a BHE cross-section served as a benchmark test, where the OGS implementation was compared against an independent FreeFem++ implementation. The results seemed identical.
Considering the H+freezing sub-problem, we showed a proof of concept. The simulation of an emerging ice barrier led to the expected feedback on the hydraulic flow.
Considering the M+freezing sub-problem, we first verified that a phase change from water to ice generates exactly the assumed $9$\% volumetric expansion. Then, effects of a more sophisticated thermo-mechanical loading path %prescribing the $u,T$ signal
on a freezing/thawing poro-elastic column was studied. The obtained reaction forces seemed plausible, thus yielding another proof of concept.
All in all, this work presents a suite of benchmark examples suitable for testing, verifying and comparing THM+F implementations. As a next step, the multi-field couplings TH+F, TM+F and finally THM+F will be tested, verified and compared. Challenges are expected such as the freezing front localization and the highly nonlinear freezing feedback on the hydraulic flow.

\begin{figure}[h!]
\begin{center}
\includegraphics[width=1.0\textwidth]{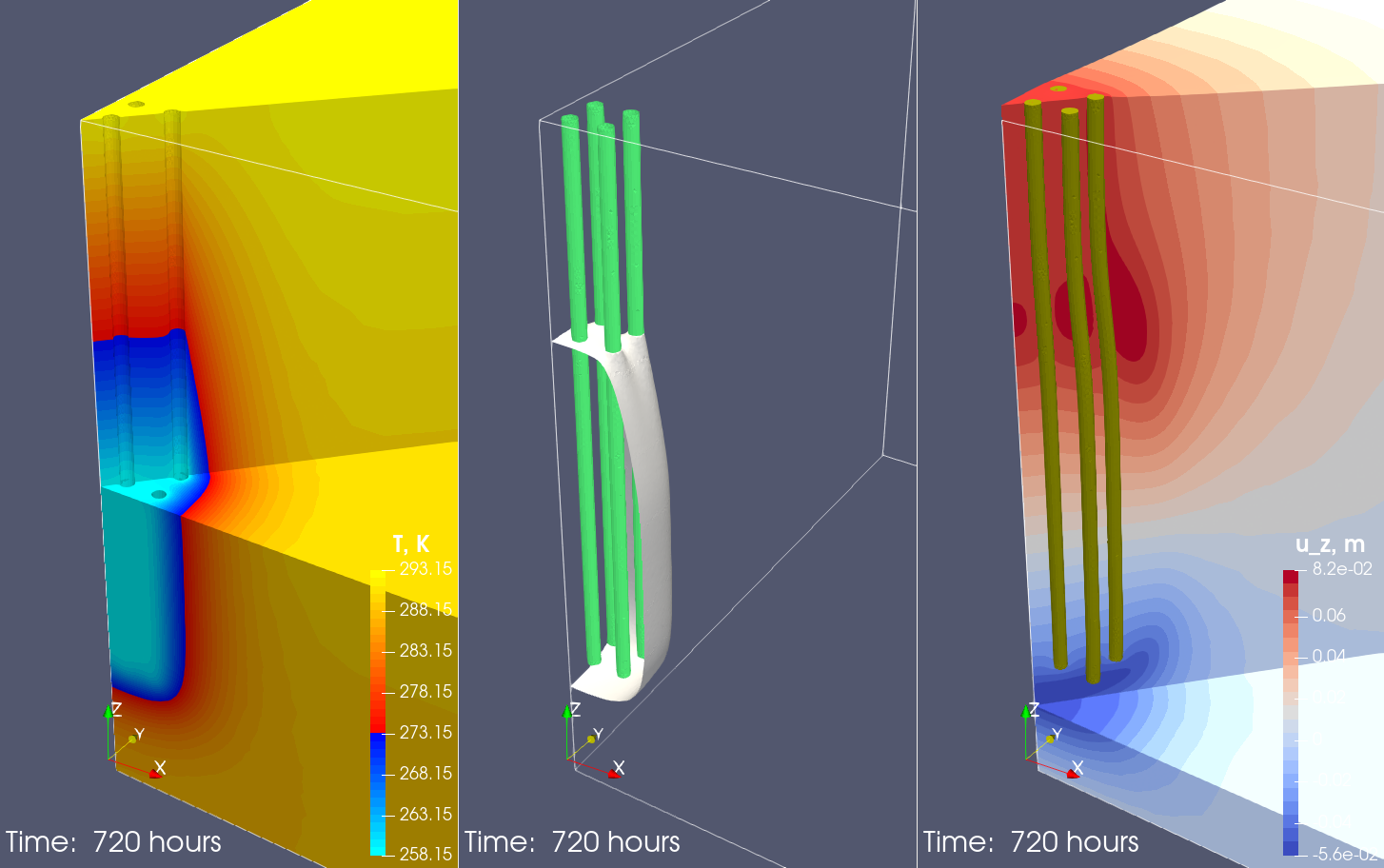} 
\end{center}
\caption{Temperature distribution, ice volume fraction and deformed configuration (the vertical displacement component) of soil specimen in the vicinity of BHEs after 720 hours of cooling.}
\label{fig_GEWS2d_4}
\end{figure}

To finalize the paper, we get back to the geothermal application from the Introduction and give a short preview of the simulation results of a ground-based energy storage system (GEWS) depicted in \autoref{fig_GEWS_project}. Due to the symmetry of the original block of soil with a group of 16 BHEs, computational domain considered consists of just a quarter of the block. Spatial and temporal evolution of coolant temperature inside of each BHE which serves as thermal input that triggers soil freezing around BHEs is given by equation (\ref{T1onGammaD}) and \autoref{fig_9}. More generally, all dimensions, time-scale and material parameters relevant to the corresponding experimental setup can be found in Section \ref{T+freezing_SoilFr} where a simpler (2D, TH) model has already been tested. Here, we compute the process taking 720 hours (30 days) of cooling -- after 360 hours, it is essentially a steady-state process. The left plot in \autoref{fig_GEWS2d_4} depicts the temperature evolution in the soil. The color legend is chosen such that one can easily distinguish both ice and liquid fractions. The mid plot specifies the ice volume fraction explicitly. Finally, the right one presents the vertical component of soil displacement plotted in a deformed configuration. One can particularly observe how the BHEs deform due to the frozen soil deformation.

The computed results are already qualitatively intuitive. Part II will present a complete detailed analysis of the outlined field experiment, including site-specific parameter calibration of the THM+freezing model, realistic aquifer geometry, and flow description. Numerical aspects and potential computational challenges will be also dissected.

%%%%%%%%%%%%%%%%%%%%%%%%%%%%%%%%%%%%%%%%%%%%%%%%%%%%%%%%%%%%%%%%%%%%%%%%%%%%%%%%%%%%%%%%%%%%%%%%%%%%%%%%%%%%%%%%%%
\newpage
\section{Appendix}

\paragraph{On the choice of exponent in the sigmoid function $S_\mathrm{I}$}
Parameter $k$ defines the steepness of the function $S_\mathrm{I}$ around the value $T=T_\mathrm{fr}$ or, in other words, the thickness (or width) of what can be called a \emph{transition zone} of $S_\mathrm{I}$. How can $k$ be chosen? First, since 1 to 0 are only the limiting values of $S_\mathrm{I}$ of in the infinite interval $(-\infty,\infty)$, we want to clarify a notion of transition zone. Let $\epsilon\ll 1$ be a user defined-parameter (a practical magnitude can be, e.g., 0.01) and $\Delta>0$ such that $S_\mathrm{I}(T_\mathrm{fr}-\Delta)=1-\epsilon$ and $S_\mathrm{I}(T_\mathrm{fr}+\Delta)=\epsilon$. We call the interval $(T_\mathrm{fr}-\Delta,T_\mathrm{fr}+\Delta)$ on which function $S_\mathrm{I}$ transits from $1-\epsilon$ to $\epsilon$ a transition zone and its length of $2\Delta$ will then be a thickness/width of a transition zone. Then, the relation between $k$ and $\Delta$ is equivalently derived from either of the above equations. Thus, we obtain $k=\Delta^{-1}\ln (\epsilon^{-1}-1)$. This is not a final result for $k$ yet since $\Delta$ remains undefined. To choose $\Delta$ we require that $(T_\mathrm{fr}-\Delta,T_\mathrm{fr}+\Delta)$ must be naturally embedded into the range $(T_\mathrm{lower},T_\mathrm{upper})$ with $T_\mathrm{lower}<T_\mathrm{fr}$ and $T_\mathrm{upper}>T_\mathrm{fr}$ being prescribed by the particular problem at hand, and, furthermore, one must ensure that the ratio $\frac{T_\mathrm{upper}-T_\mathrm{lower}}{2\Delta}$ is larger than 1.

The ultimate result reads as follows: given $T_\mathrm{fr}$, $T_\mathrm{lower}$ and $T_\mathrm{upper}$, one fixes 
\begin{equation}
    \Delta<\min\Bigl\{T_\mathrm{upper}-T_\mathrm{fr},T_\mathrm{fr}-T_\mathrm{lower},\frac{1}{2}(T_\mathrm{upper}-T_\mathrm{lower})\Bigr\}.
\label{Delta-k-1}
\end{equation}
With the prescribed $\epsilon\ll 1$, the desired representation for $k$ then follows:
\begin{equation}
 k=\Delta^{-1}\ln (\epsilon^{-1}-1).
\label{Delta-k-2}
\end{equation}

Notice that in the computational results presented in the manuscript, the above criterion is always fulfilled/ensured even though we do not show this explicitly. 

%%%%%%%%%%%%%%%%%%%%%%%%%%%%%%
\paragraph{On the convexity of energy functional $E(T)$ to enable a convergent iterative minimization/solution process}
The Sigmoid function $S_\mathrm{I}$ present in the heat conduction equation (\ref{BalEn}) makes it (strongly) non-linear in the unknown $T$ to be solved for. One of the computational issues that may occur while solving the corresponding weak formulation using, e.\,g., the Newton-Raphson method is the convergence behavior of this iterative process. The apparent reason is that $S_\mathrm{I}$ also leads to non-convexity of the weak problem, more precisely, of the underlying energy functional. Below, we illustrate this issue and propose one of the remedies in coping with it.

In Eq. (\ref{BalEn}), for the sake of simplicity, we set $Q_T=0$ and neglect the advection term, thus assuming no coupling between hydraulic flow and temperature. Let us also denote
\begin{equation*}
    (\varrho c_p)_1:=
    (1-\phi)\varrho_\mathrm{SR}c_{p\mathrm{S}}+\phi\varrho_\mathrm{LR}c_{p\mathrm{L}}, \qquad (\varrho c_p)_2:=-\phi\varrho_\mathrm{LR}c_{p\mathrm{L}}+\phi\varrho_\mathrm{IR}c_{p\mathrm{I}},
\end{equation*}
and
\begin{equation*}
    \lambda_1:=
    (1-\phi)\lambda_\mathrm{SR}+\phi\lambda_\mathrm{LR}, \qquad \lambda_2:=-\phi\lambda_\mathrm{LR}+\phi\lambda_\mathrm{IR}.
\end{equation*}
Rewriting the corresponding terms in Eq. (\ref{BalEn}) as
\begin{equation*}
    (\varrho c_p)^\mathrm{eff}=(\varrho c_p)_1+(\varrho c_p)_2 S_\mathrm{I}(T) \quad \mbox{and} \quad
    \lambda^\mathrm{eff}=
    \lambda_1+\lambda_2 S_\mathrm{I}(T),
\end{equation*}
the considered equation converts into
\begin{equation}
    \bigl[
    (\varrho c_p)_1+(\varrho c_p)_2 S_\mathrm{I}(T)-\ell\varrho_\mathrm{IR}\phi S_\mathrm{I}^\prime(T)
    \bigr] \frac{\partial T}{\partial t}
    -\mathrm{div}
    \bigl[ (\lambda_1+\lambda_2 S_\mathrm{I}(T)) \nabla T \bigr] = 0.
\label{BalEn_C}
\end{equation}
Within the implicit time-discretizations treatment, the related weak formulation reads: 
At any time step $n$, find $T^n$ belonging to admissible space\footnote{Theory which studies the solvability of PDEs and the corresponding solution regularity (differentiability) properties defines an admissible space for ... as $H^1$-Sobolev space (another standard notation is $W^{1,2}$) } such that
\begin{align}
a(T^n,v):=\int_\Omega 
& \Bigl\{  \bigl[
    (\varrho c_p)_1+(\varrho c_p)_2 S_\mathrm{I}(T^n)-\ell\varrho_\mathrm{IR}\phi S_\mathrm{I}^\prime(T^n)
    \bigr](T^n-T^{n-1})v \Bigr. \notag\\
& \Bigl. -\Delta t \bigl[\lambda_1+\lambda_2 S_\mathrm{I}(T)\bigr] \nabla T^n \cdot\nabla v
\Bigr\}\,\mathrm{d}{\bm x}=0
\label{a(T,v)}
\end{align}
holds for all test functions $v$ from the test space. Note that $T^{n-1}$ is known from the previous time step, $\Delta t$ is a time-step increment, and $\mathrm{d}\bm{x} \in \partial\Omega$ is the tuple of coordinate differentials.
Using this and the calculus of variation, the energy functional $E$, whose variation yields the above $a(T^n,v)=0$, can be derived. We get
\begin{align}
E(T^n):= \int_\Omega (\varrho c_p)_1 & \tfrac{1}{2}(T^n-T^{n-1})^2 \mathrm{d}{\bm x} \nonumber \\
+ \int_\Omega (\varrho c_p)_2 & \Bigl[ \tfrac{1}{2}(T^n-T_\mathrm{fr})(T^n+T_\mathrm{fr}-2T^{n-1})-\tfrac{1}{k^2}\,\mathrm{dilog}(S_\mathrm{I}^{-1}(T^n)) \nonumber \\
   & +\tfrac{1}{k}(T^n-T^{n-1})\ln(S_\mathrm{I}(T^n)) \Bigr]\mathrm{d}{\bm x} \nonumber \\
+ \int_\Omega \ell\varrho_\mathrm{IR}\phi & \Bigl[ -(T^n-T^{n-1})S_\mathrm{I}(T^n)
     + T^n-T_\mathrm{fr}+\tfrac{1}{k}\ln(S_\mathrm{I}(T^n)) \Bigr]\mathrm{d}{\bm x} \nonumber \\
- \int_\Omega \Delta t & \bigl[\lambda_1+\lambda_2 S_\mathrm{I}(T^n)\bigr] |\nabla T^n|^2\mathrm{d}{\bm x},
\label{E(T)}
\end{align}
where a so-called dilogarithm function is defined as $\mathrm{dilog}(x):=\int_1^x \ln(t)/(1-t)\mathrm{d}t$. Note that for the recovery of the last (gradient) term in $E$ we used a slight abuse of math strictness, assuming that the weak term $S_\mathrm{I}(T) \nabla T^n \cdot\nabla v$ can form the variation of $S_\mathrm{I}(T) |\nabla T^n|^2$, what actually only holds true if one replaces $S_\mathrm{I}(T)$ by $H_\mathrm{I}(T)$.

\begin{figure}[h!]
\begin{center}
\includegraphics[width=1.0\textwidth]{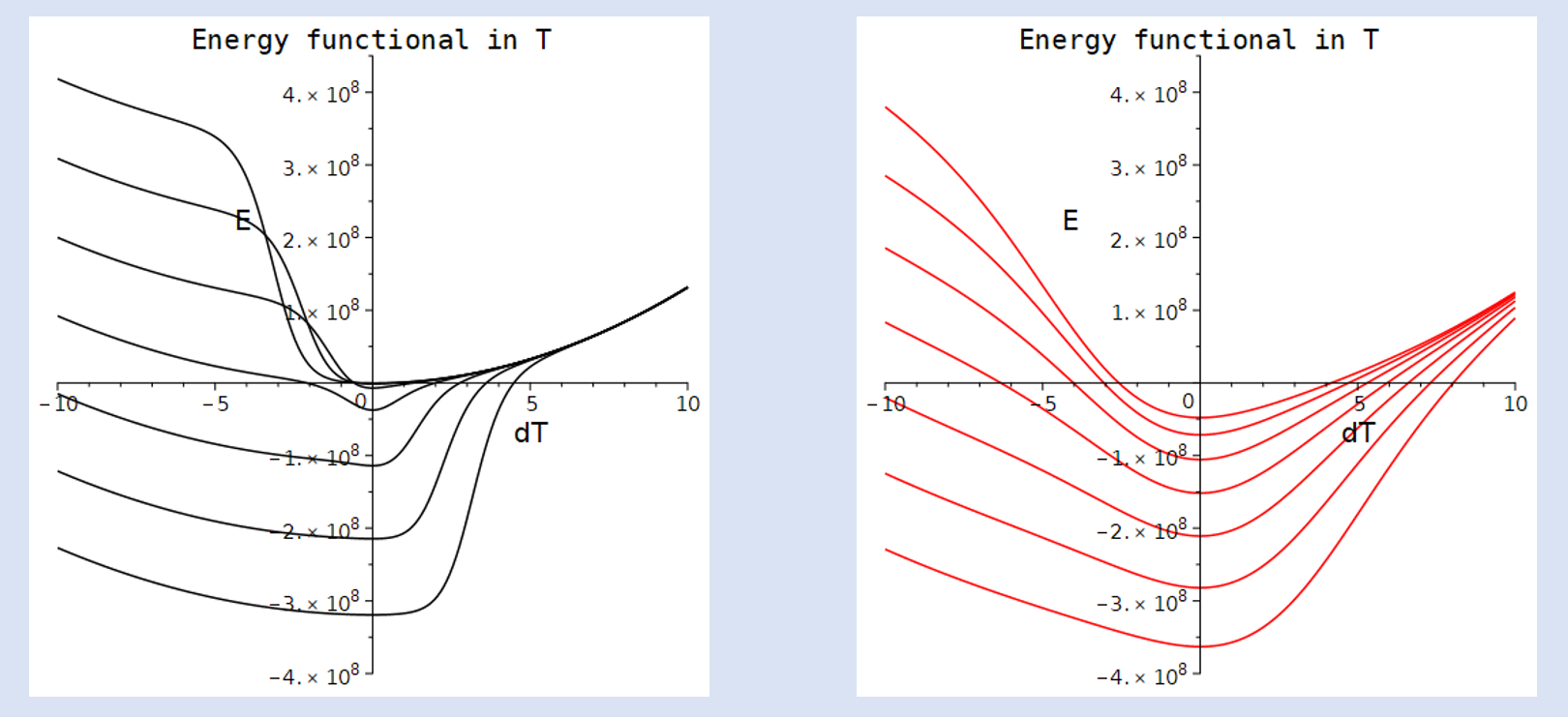}
\end{center}
\caption{$k$-parametric convexity of the energy functional $\widetilde{E}(\Delta T)$ represented by its integrand $f(\Delta T)$: for $k=2$ (on the left) $f$ is visibly non-convex, whereas the passage to $k=0.5$ (on the right) convexifies $f$ (corresponding to less steep $S_\mathrm{I}$ at $T=T_\mathrm{fr}$).}
\label{fig_E(T)_convexification}
\end{figure}

A way to quickly assess convexity of a functional is to visualize and check the convexity of the corresponding integrands with respect to their arguments. In our case, the functional is of the from $E(s(x))=\int_\Omega \bigl( f(s)+g(\nabla s \bigr) \mathrm{d}x$ and it can be seen that convexity of the integrand $f$ will be $k$-dependent. To illustrate this explicitly, we first convert $E$ in (\ref{E(T)}) into the incremental-based representation $E(T^n)=E(T^{n-1}+\Delta T)=\widetilde{E}(\Delta T)$ with $\Delta T$ being the unknown. Then $f$, the integrand of $\widetilde{E}(\Delta T)$, which in our case explicitly reads
\begin{align}
    f(\Delta T):=(\varrho c_p)_1 & \tfrac{1}{2}(\Delta T)^2 \nonumber \\
+ (\varrho c_p)_2 & \Bigl[ -\tfrac{1}{2}(T^{n-1}+\Delta T-T_\mathrm{fr})(T^{n-1}-\Delta T-T_\mathrm{fr}) \nonumber \\
   & -\tfrac{1}{k^2}\mathrm{dilog}(S_\mathrm{I}^{-1}(T^{n-1}+\Delta T))
   +\tfrac{1}{k}(\Delta T)\ln(S_\mathrm{I}(T^{n-1}+\Delta T)) \Bigr] \nonumber \\
+ \ell\varrho_\mathrm{IR}\phi & \Bigl[ -(\Delta T)S_\mathrm{I}(T^n)
     + T^{n-1}+\Delta T-T_\mathrm{fr}+\tfrac{1}{k}\ln(S_\mathrm{I}(T^{n-1}+\Delta T)) \Bigr],
\label{f(dT)}
\end{align}
can be visualized: \autoref{fig_E(T)_convexification} depicts the plots of $f(\Delta T)$ using the material data parameters from Table \ref{table_MatData_Tfreezing} for two cases of $k$. It can be seen that passage from $k=2$ to $k=0.5$ convexifies $f$.

The above consideration provides basic evidence that convexity of $E$, whose variation yields the weak and strong forms of the heat conduction problem with phase change (T+freezing problem), depends on parameter $k$. In the numerical computations, to ensure the convergence of iterative solution process for (\ref{a(T,v)}) one must pick up $k$ which fulfills both the criterion (\ref{Delta-k-1})--(\ref{Delta-k-2}), and the convexity property of $f$ in (\ref{f(dT)}). This already applies regardless of whether we solve only the heat conduction equation alone with phase change, or the coupled problem (TH, TM, THM) where the T-component is involved in. %In the latter case, the divergence is more pronounced.

%%%%%%%%%%%%%%%%%%%%%%%%%%%%%%%%%%%%%%%%%%%%%%%%%%%%%%%%%%%%%%%%%%%%%%%%%%%%%%%%%%%%%%%%%%%%%%%%%%%%%%%%%%%%%%%%%%


\newpage

\begin{thebibliography}{99}
\bibitem{Mikkola2001} M. Mikkola, and J. Hartikainen. Mathematical model of soil freezing and its numerical implementation. IJNME, 52:543–-557, 2001.

\bibitem{Graf2008} T. Graf. Multiphasic Flow Processes in Deformable Porous Media under Consideration of Fluid Phase Transitions. PhD thesis, Universit\"at Stuttgart, 2008.

\bibitem{Bluhm2011} J. Bluhm, T. Ricken, and M. Bloßfeld. Ice formation in porous media. In B. Markert et al.\ (eds.): Advances in Extended and Multifield Theories for Continua. Lecture Notes in Applied and Computational Mechanics, pp.\ 173--174. Springer, Berlin, 2011.

%%%%%%%%%%%%

\bibitem{Luckner1989} L. Luckner, M.Th. van Genutchen, and D.R. Nielsen. A consistent set of parametric models for the two-phase flow of immiscible fluids in the subsurface. Water Resour. Res.,  25(10):2187–-2193, 1989.

\bibitem{Grant2000} S. Grant. Physical and chemical factors affecting contaminant hydrology in cold environment. ERDC/CRREL TR-00-21, US Army Corps of Engineers, 2000.

\bibitem{Coussy2005} O. Coussy. Poromechanics of freezing materials. J. Mech. Phys. Solids, 53:1389–-1718, 2005.

\bibitem{Nishimura2009} S. Nishimura, A. Gens, S. Olivella, and R.J. Jardine. THM-coupled finite element analysis of frozen soil: formulation and application. Geotechnique, 59(3):159—171, 2009.

\bibitem{Zhou2014} M.M. Zhou. Computational simulation of soil freezing: multiphase modeling and strength upscaling. Ph.D. thesis. Ruhr University Bochum, 2014.

\bibitem{NaSun2017} S.H. Na, and  W.C. Sun. Computational thermo-hydro-mechanics for multiphase freezing and thawing porous media in the finite deformation range. Comput. Methods Appl. Mech. Engrg., 318:667--700, 2017.

\bibitem{Bekele2017} Y. Bekele, H. Kyokawa, A. M. Kvarving, T. Kvamsdal, S. Nordal. Physics,  Isogeometric analysis of THM coupled processes in ground freezing. Computers and Geotechnics, 88:129--145, 2017.

%%%%%%%%%%%%

\bibitem{Sweidan2020} A.H. Sweidan, Y. Heider, and B. Markert. A unified water/ice kinematics approach for phase-field thermo-hydro-mechanical modeling of frost action in porous media. Comput. Methods Appl. Mech. Engrg., 372:113358, 2020.

\bibitem{Sweidan2021} A.H. Sweidan, K. Niggemann, Y. Heider, M. Ziegler, and B. Markert. Experimental study and numerical modeling of the therm-hydro-mechanical processes in soil freezing with different frost penetration directions. Acta Geotechnica, 17:231-255, 2022.

\bibitem{AlexSol} V. Alexiades and A.D. Solomon. Mathematical Modeling of Melting and Freezing Processes. Hemisphere Publishing Corporation, Washington, 1993.

\bibitem{FreeFem} F. Hecht, A. Leharic, and O. Pironneau. FreeFem++: Language for finite element method and Partial Differential Equations (PDE), Universit\'e Pierre et Marie, Laboratoire Jacques-Louis Lions, http://www.freefem.org/ff++/.

\bibitem{GPL2020} Geothermiepreisliste 2020, http://www.frank-gmbh.de.

\bibitem{Andersland2004} O.B. Andersland and B. Ladanyi. Frozen Ground Engineering. 2nd ed. John Wiley \& Sons, Chichester, 2004.

\bibitem{Jessberger1980} H.L. Jessberger. Theory and application of ground freezing in civil engineering. Cold Regions Science and Technology, 3:3--27, 1980.

\bibitem{French2007} H.M. French. The Periglacial Environment. 3rd ed. John Wiley \& Sons, Chichester, 2007.

\bibitem{Riseborough2008} D. Riseborough, N. Shiklomanov, B. Etzelm{\"u}ller, S. Gruber, and S. Marchenko. Recent advances in permafrost modelling. Permafrost and Periglacial Processes, 19(2):137--156, 2008.

\bibitem{Stefan1891} J. Stefan. {\"U}ber die Theorie der Eisbildung, insbesondere {\"u}ber die Eisbildung im Polarmeer. Annalen der Physik und Chemie, 42:269--286, 1891.

\bibitem{Alexiades1993} V. Alexiades and A.D. Solomon. Mathematical Modeling of Melting and Freezing Processes. Hemisphere Publishing Corporation, Washington, D.C., 1993.

\bibitem{Biot1941} M.A. Biot. General Theory of Three-Dimensional Consolidation. Journal of Applied Physics, 12(2):155--164, 1941.

\bibitem{deBoer2000} R. de Boer. Theory of Porous Media: Highlights in Historical Development and Current State. Springer, Berlin, 2000.

\bibitem{Ehlers2002} W. Ehlers and J. Bluhm (eds.). Porous Media: Theory, Experiments and Numerical Applications. Springer, Berlin, 2002.

\bibitem{Voller1987} V.R. Voller and C. Prakash. A fixed grid numerical modelling methodology for convection-diffusion mushy region phase-change problems. International Journal of Heat and Mass Transfer, 30(8):1709--1719, 1987.

\bibitem{Kolditz2012} O. Kolditz, S. Bauer, L. Bilke, N. B{\"o}ttcher, J.O. Delfs, T. Fischer, U.J. G{\"o}rke, T. Kalbacher, G. Kosakowski, C.I. McDermott, C.H. Park, F. Radu, K. Rink, H. Shao, H.B. Shao, F. Sun, Y.Y. Sun, A.K. Singh, J. Taron, M. Walther, W. Wang, N. Watanabe, Y. Wu, M. Xie, W. Xu, and B. Zehner. OpenGeoSys: an open-source initiative for numerical simulation of thermo-hydro-mechanical/chemical (THM/C) processes in porous media. Environmental Earth Sciences, 67(2):589--599, 2012.

\bibitem{Bilke2019} L. Bilke, B. Flemisch, T. Kalbacher, O. Kolditz, R. Helmig, and T. Nagel. Development of Open-Source Porous Media Simulators: Principles and Experiences. Transport in Porous Media, 130(1):337--361, 2019.


\end{thebibliography}
\end{document}